\documentclass[review]{elsarticle}
\pdfoutput=1
\usepackage{lineno,hyperref}
\usepackage{graphicx}
\usepackage[linesnumbered,ruled,lined]{algorithm2e}
\usepackage{float}
\usepackage{tikz}
\usepackage{epstopdf}
\usepackage{amssymb}
\usepackage{amsmath}
\usepackage{subcaption}
\usepackage{footnote}
\usepackage{array}
\usepackage{cases}
\usepackage{multirow}
\usepackage{color}
\usepackage{hyperref}
\graphicspath{ {./Figure/} }
{\color{red}\journal{Elsevier}}
\begin{document}
%%%%%%%%%%%%%%%%%%%%%%%%%%%%%%%%%%%%%%%%%%%%%%%%%%%%%%%%%%%%%
%
% Section
%
%%%%%%%%%%%%%%%%%%%%%%%%%%%%%%%%%%%%%%%%%%%%%%%%%%%%%%%%%%%%%	
\begin{frontmatter}
		\title{Physics-driven complex relaxation for multi-body systems of SPH method}
		\author[myfirstaddress]{Chenxi Zhao}
		\ead{chenxi.zhao@tum.de}
		\author[myfirstaddress]{Yongchuan Yu}
		\ead{yongchuan.yu@tum.de}
		\author[myfirstaddress]{Oskar J. Haidn}
		\ead{oskar.haidn@tum.de}
		\author[mysecondaryaddress]{Xiangyu Hu \corref{mycorrespondingauthor}}
		\ead{xiangyu.hu@tum.de}
		\address[myfirstaddress]{Chair of Space Propulsion and Mobility, 
			Technical University of Munich, 85521 Ottobrunn, Germany}
		\address[mysecondaryaddress]{Chair of Aerodynamics and Fluid Mechanics, 
			Technical University of Munich, 85748 Garching, Germany}
		\cortext[mycorrespondingauthor]{Corresponding author. }

\begin{abstract}
In the smoothed particle dynamics (SPH) method, the characteristics of a target particle are 
interpolated based on the information from its neighboring particles. Consequently, a uniform 
initial distribution of particles significantly enhances the accuracy of SPH calculations. 
This aspect is particularly critical in Eulerian SPH, where particles are stationary throughout 
the simulation. To address this, we introduce a physics-driven complex relaxation method for 
multi-body systems. Through a series of two-dimensional and three-dimensional case studies, 
we demonstrate that this method is capable of achieving a globally uniform particle distribution, 
especially at the interfaces between contacting bodies, and ensuring improved zero-order 
consistency. Moreover, the effectiveness and reliability of the complex relaxation method in 
enhancing the accuracy of physical simulations are further validated.
\end{abstract}


\end{frontmatter}
%%%%%%%%%%%%%%%%%%%%%%%%%%%%%%%%%%%%%%%%%%%%%%%%%%%%%%%%%%%%%
%
% Section
%
%%%%%%%%%%%%%%%%%%%%%%%%%%%%%%%%%%%%%%%%%%%%%%%%%%%%%%%%%%%%%
\section{Introduction} \label{sec:introduction}
As a meshless computational approach, the smoothed particle hydrodynamics (SPH) method discretizes the 
continuum media with a set of particles and employs a kernel function interpolation for the 
calculation of interactions among neighboring particles. Initially proposed by Lucy 
\cite{lucy1977numerical} and by Gingold and Monaghan \cite{gingold1977smoothed} in 1977 for 
astrophysical applications, the SPH method has been implemented in fluid dynamics 
\cite{monaghan1994simulating, shao2012improved, ye2019smoothed}, solid mechanics 
\cite{benz1995simulations, wu2023essentially}, fluid-structure interaction 
\cite{amicarelli2020sphera, zhang2021multi} and heat transfer \cite{tang2023integrative}. Its 
practical applications have been successfully extended to diverse areas, including ocean engineering 
\cite{zhang2017smoothed, luo2021particle}, aerospace engineering \cite{calderon2019sph}, civil 
engineering \cite{bui2013improved}, biomechanics \cite{zhang2023multi}, and laser welding 
\cite{russell2018numerical}. 

Although it has many successful applications in different fields, the SPH method still challenges 
important improvements in several aspects. Getting a body-fitted and isotropic initial 
particle distribution \cite{colagrossi2012particle, vacondio2021grand,
pourlak2023importance} is one of the critical issues that need to be solved. 
In the early days of the development of the SPH method, there were typically two main categories 
of generating particle distributions: generating particles at the center of a spatially divided 
lattice unit \cite{dominguez2011development} or in the center of tetra- or hexahedron volume 
elements \cite{vignjevic2013parametric, heimbs2011computational}. The volumetric-mesh converted 
particle distributions are adept at representing complex surfaces but often face challenges 
in achieving uniform spacing and volume. In comparison, cubic lattice-based particle 
initiation promotes a more homogeneous and isotropic distribution, but it's hard to get a 
body-fitted distribution. Further advancements focus on resettling particles towards 
body-fitted, especially for geometries with complicated features like sharp corners, narrow gaps, 
and irregular surfaces. 
Colagrossi \cite{colagrossi2012particle} introduced kernel gradient force and damping force in the 
momentum equation of their particle packing algorithm to provide an `equilibrium' initial particle 
distribution for basic geometries. Jiang et al. \cite{jiang2015blue} applied an additional 
cohesive force proposed by Akinci et al. \cite{akinci2013coupling} to adjust 
the density difference between surface points and interior points in addition to the Kernel gradient 
force and the surface force. This approach concentrates on packing particles both within the geometry 
and on its surface, as well as prevents the generation of particles outside the boundary.
The ALARIC method \cite{vela2018alaric} solves the SPH form dynamic equations 
by introducing external forces and damping forces. With its unique sensible criteria for nullifying
velocity, it can achieve arbitrarily complex density distributions for particles in order to 
obtain isotropic initial conditions without preferred directions of propagation.
Furthermore, Litvinov \cite{litvinov2015towards} pointed out that with constant background pressure
adopted by a transport-velocity formulation in weakly compressible SPH (WCSPH) simulation can achieve 
a self-relaxation mechanism, thus resulting in the zero-order consistency of particle distribution,
as well as the same high-order integration error as the uniform grid could achieve.
Based on this theory, Fu \cite{fu2019optimal, 
fu2019isotropic} supplemented the equation of state (EOS) to achieve a physics-motivated relaxation 
process, thus obtaining the isotropic distribution of particles with adaptive smoothing length. 
Ji further \cite{ji2021feature} substituted the additional ghost particles with a feature 
boundary correction term, addressing incomplete kernel support at boundaries and reducing 
computational costs. 
Zhu et al. \cite{zhu2021cad} went on to streamline the physics-driven relaxation method with a 
surface bounding algorithm based on a level-set field, achieving a body-fitted and isotropic particle
distribution for complex geometry. By replacing the surface bounding method with the 
"static confinement" boundary condition to complete kernel support of boundary particles, 
Yu et al. \cite{yu2023level} improves the zero-order consistency of the physics-driven 
relaxation method at the boundaries of arbitrary complex geometries.

The aforementioned particle generation methods primarily focus on single-body systems, 
and most of them are aimed at the solid boundaries in a simulation system. 
This is due to the fact that in the Lagrangian framework, fluid particles shift from 
their initial positions to achieve a body-fitted and isotropic distribution as the simulation
goes on via a so-called "self-adjustment" mechanism. Even so, those particles with 
“in-equilibrium” initial positions can induce spurious motions upon redistribution, 
which strongly affect the fluid evolution \cite{colagrossi2012particle}.
Moreover, the distribution of particles across multi-body systems, encompassing both 
fluids and single or numerous solids, holds greater significance for the Eulerian formulation 
of the SPH method \cite{Eulerian-SPH-lind2016high, Eulerian-SPH-nasar2016eulerian, wang2023eulerian}, 
which has seen enlarging applications recently due 
to its higher spatial accuracy and increased computational efficiency.
This is because, under the Eulerian system, fluid particles do not implement the `self-adjustment' 
mechanism to rearrange their positions to an equilibrium state. 
Therefore, whether the initial particle position in the Eulerian method, especially the 
particles at the multi-body interface, 
can guarantee zero-order consistency becomes more crucial.
However, it is a difficult challenge to ensure that the initial particles obtain a 
body-fitted isotropic distribution in a multi-body system, and even achieve zero-order 
consistency at the interface. For simple geometries, particle arrangement can be 
programmed by manually calculating the positions of surface particles 
\cite{wang2021graphics, huang2019kernel}. Shadloo \cite{shadloo2011improved} calculates 
the discrete points of the two-dimensional airfoil surface through formulas and generates 
boundary particles at the corresponding positions. Cooperating with a complex multiple 
boundary tangents (MBT) method is used to reorganize boundary particles and their adjacent 
fluid particles to ensure the uniformity of particle distribution. However, 
no concrete parameters for particle distribution near the boundary in this method are given for estimate.
Negi \cite{negi2021algorithms} advanced this field by proposing a technique that merges the strategies 
from Colagrossi et al. \cite{colagrossi2012particle} and Jiang et al. \cite{jiang2015blue}, 
rearranging particles across both internal and external domains simultaneously. 
This approach employs kernel gradient forces for particle movement, 
introducing a repulsive force when the distance between two particles falls below 
the particle spacing. Surface particles are identified based on their proximity to 
boundary nodes, and interior and exterior particles are detected at the end of the simulation. 
This method does not achieve zero-order consistency successfully at the interface of two bodies, 
resulting in a distribution that cannot be considered uniformly isotropic.

\begin{figure}
  \centering
  \includegraphics[width=4in]{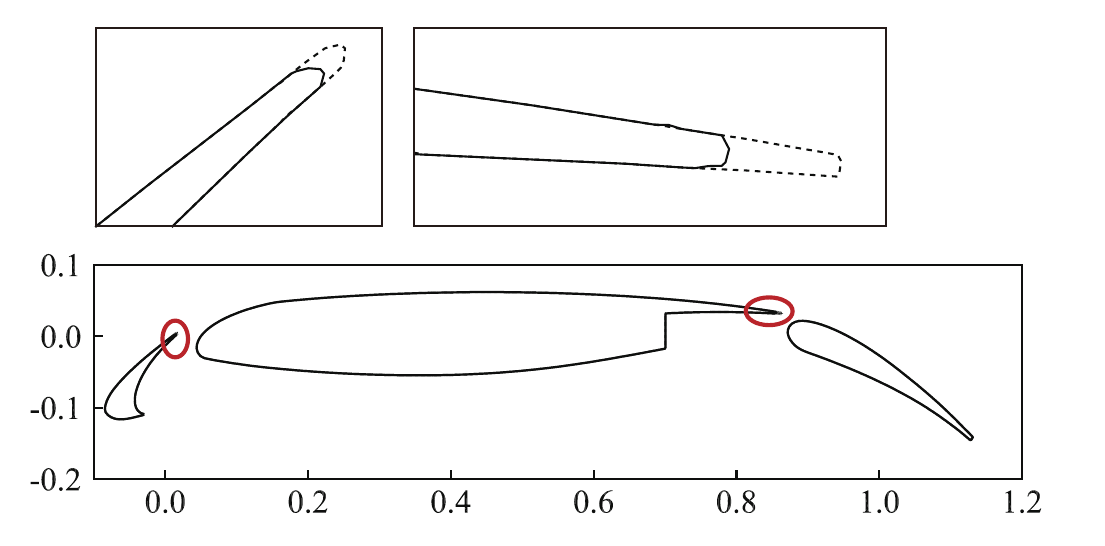}
  \caption{Geometric boundary analysis after importing airfoil and its external flow field 
  with the same boundary into the SPH solver. In the CAD stage, the airfoil and its external 
  fluid field shares the same boundary. Although the same higher resolution $\Delta x = L/1000$ 
  is used for discretization, due to the discrete origin points of the airfoil (the leading edge) 
  and the discrete origin point of the external flow field 
  (lower left corner of the calculation domain ) are different, 
  resulting in a small gap at the trailing edge of the airfoil between the solid body and the fluid body.
  Note that the solid line is the geometric boundary of the inner solid body, 
  while the dashed line is that of the outer fluid body.}
  \label{fig: airfoil-levelset-profile}
\end{figure}

Meanwhile, with the in-depth application of the SPH method in various industrial fields, 
complex geometric shapes are included in the discussion of SPH simulation. 
Usually, these complex geometries are generated by CAD software and decomposed into 
the SPH solver through a specific parser. For different geometries, different parsers will 
set discrete origins according to the outer contour of the geometry while parsing into 
the SPH solver. However, for relatively coarse resolutions, the outer contour of the 
geometry may not be fully resolved. Therefore, even if the boundaries are exact, 
different outer contour dimensions and discrete origin settings may cause gaps 
between multi-body systems with the same boundaries, thus resulting in inconsistent topological 
structures and affecting the simulation accuracy. 
(as shown in Fig.\ref{fig: airfoil-levelset-profile}).
In the Lagrangian form of the SPH method, these gaps between the fluid domain and 
the solid domain do not seem to have much impact. Since the fluid particles could re-arrange the 
position during the "self-adjustment" mechanism. However, for Eulerian form sph amplifiers, 
these gaps may even cause the calculation to crash.

On this basis, this paper first adopts a simple method of performing Boolean operations based on 
level-set fields to ensure that there will be no gaps between interfaces in multi-body systems. 
This method does not depend on the specific parser or the specific geometry format. 
Then, we introduced a method for particle generation in multi-body systems, 
which is named as the complex relaxation process in this manuscript. 
The complex relaxation method can not only ensure that different bodies obtain 
body-fitted and isotropic particle distributions but also ensure the zero-order consistency 
of the particle distribution at the interface of the multi-body system, 
especially the particle distribution on the fluid side.
In addition, this algorithm does not need to arrange boundary nodes in advance, 
nor does it need to introduce ghost particles to fill the kernel function support domain, 
so the calculation efficiency can reach a very high level.

The framework of this work is as follows: Section \ref{section: methodology} details the process of 
implementing the complex relaxation, including relevant formulas and their SPH discretization. 
Section \ref{section: relaxation-cases} displays the initial particle distributions achieved in both 
two and three dimensions through our algorithms. In Section \ref{section: physical-cases}, we 
present the simulation results of a physical case using particle distribution generated by our 
method. The paper concludes with a summary of our findings in the final section. The research 
presented herein was conducted using the SPHinXsys platform \cite{zhang2021sphinxsys}.

%%%%%%%%%%%%%%%%%%%%%%%%%%%%%%%%%%%%%%%%%%%%%%%%%%%%%%%%%%%%%
%
% Section
%
%%%%%%%%%%%%%%%%%%%%%%%%%%%%%%%%%%%%%%%%%%%%%%%%%%%%%%%%%%%%%
\section{Methodology} \label{section: methodology}

This section first provides the physics-driven relaxation algorithm, along with brief descriptions of 
boundary processing methods completed previously, including surface bounding and static confinement. 
Following this, the detailed procedure for multi-body relaxation referring to geometry parse and particle 
generation is elaborately presented.
%%%%%%%%%%%%%%%%%%%%%%%%%%%%%%%%%%%%%%%%%%%%%%%%%%%%%%%%%%%%%
% Section
%%%%%%%%%%%%%%%%%%%%%%%%%%%%%%%%%%%%%%%%%%%%%%%%%%%%%%%%%%%%%
\subsection{Preliminary work}
%%%%%%%%%%%%%%%%%%%%%%%%%%%%%%%%%%%%%%%%%%%%%%%%%%%%%%%%%%%%%
% Section
%%%%%%%%%%%%%%%%%%%%%%%%%%%%%%%%%%%%%%%%%%%%%%%%%%%%%%%%%%%%%
\subsubsection{Level-set method}

To accurately represent complex geometries, the level-set field $\phi(x, y, z, t)$ defined as a signed 
distance function is employed. The geometry surface $\Gamma$ can be represented by the zero level-set 
contour as
\begin{equation}\label{eq: level-set}
    \Gamma = \{(x, y) | \phi(x, y, t) = 0 \} .
\end{equation}
The normal direction $N = (n_x, n_y, n_z)^T$ of the surface can be computed from
\begin{equation}\label{eq: normal direction}
    N = \frac{\nabla \phi}{|\nabla \phi |} .
\end{equation}

This method enables a clear distinction between the inside and outside regions of the geometry. The 
exterior domain is identified as the positive phase $\phi > 0$, 
while the interior domain is represented by the region of negative phase $\phi < 0$. 
Calculate the distance from any point in the computational field to the boundary of the imported geometry
via the Simbody\cite{simbody--sherman2011simbody} or boost built-in functions, 
so that the level-set field can be established in the discrete Cartesian coordinate system.

%%%%%%%%%%%%%%%%%%%%%%%%%%%%%%%%%%%%%%%%%%%%%%%%%%%%%%%%%%%%%
% Section
%%%%%%%%%%%%%%%%%%%%%%%%%%%%%%%%%%%%%%%%%%%%%%%%%%%%%%%%%%%%%
\subsubsection{Physics-driven particle relaxation}
\label{subsec:Physics-driven-particle-relaxation} 

The foundation of the relaxation approach begins with a preconditioned Lattice particle distribution 
within the interior domain of a single geometry. 
By applying a physics-driven relaxation process governed by the conservation of momentum, 
combined with some specific boundary conditions, particles can achieve an isotropic and 
body-fitted distribution within this single geometry.
This process evolves the particles according to the following transport-velocity equation:
\begin{equation}\label{eq: transport-velocity}
    \frac{\mathrm{d} \mathbf{v}}{\mathrm{d}t} = \mathbf{F}_p .
\end{equation}
Where the $\mathbf{v}$ represents the advection velocity, and the $\mathbf{F}_p$ symbolizes the 
accelerations resulting from the repulsive pressure force. The material derivative is expressed as 
$\frac{\mathrm{d}\left(\bullet \right) }{\mathrm{d} t}=\frac{\partial \left(\bullet \right)}{\partial 
t} + \mathbf{v} \cdot \nabla \left(\bullet \right)$. In practice, the force term on the right-hand 
side of Eq.(\ref{eq: transport-velocity}) is achieved by imposing a constant background pressure 
to ensure an isotropic particle distribution as proposed by Litvinov et al. \cite{litvinov2015towards}:
\begin{equation}\label{eq: momentum-sph}
    \mathbf{F}_{p,i} = -\frac{2 p_{0} V_i}{m_i}\sum_j \nabla_i W_{ij} V_j,
\end{equation}
Here, $m$ is the particle mass, $V$ is the particle volume, $p_0$ is the constant 
background pressure, and $\nabla_i W_{ij}$ denotes the gradient of the kernel function 
$W(|\mathbf{r}_{ij}|, h)$ with respect to particle $i$. 
The terms $\mathbf{r}_{ij} = \mathbf{r}_{i} - \mathbf{r}_{j}$ and $h$ refer to the relative position 
vector and the smoothing length, respectively.

To ensure numerical stability, the time-step size $\Delta t$ is limited by the body force criterion:
\begin{equation}\label{eq: time-step}
    \Delta t \leq 0.25 \sqrt{\frac{h} {\left|\mathrm{d}\mathbf{v} / \mathrm{d}t\right|} },
\end{equation}
Subsequently, the particle positions are updated as follows:
\begin{equation}\label{eq: update-position}
    \mathbf{r^{n+1}}=\mathbf{r^n} + \mathrm{d}\mathbf{r} =\mathbf{r^n} + \frac{1}{2} \mathbf{F}_p^n \Delta t^2.
\end{equation}
Notably, at the beginning of each time step, particle velocity is reset to zero to facilitate a fully 
stationary state.

For an equilibrium state to be achieved, where particles are uniformly distributed, the kernel gradient 
summation $\sum_j \nabla_i W_{ij} V_j$ should be equal to zero to satisfy zero-order consistency. 
This situation indicates that at the end of the relaxation process, 
the forces acting between particles remain balanced, and the positions of the 
particles will not change over time.
%%%%%%%%%%%%%%%%%%%%%%%%%%%%%%%%%%%%%%%%%%%%%%%%%%%%%%%%%%%%%
% Section
%%%%%%%%%%%%%%%%%%%%%%%%%%%%%%%%%%%%%%%%%%%%%%%%%%%%%%%%%%%%%
\subsubsection{Surface bounding method}
\begin{figure}
  \centering
  \includegraphics[width=4in]{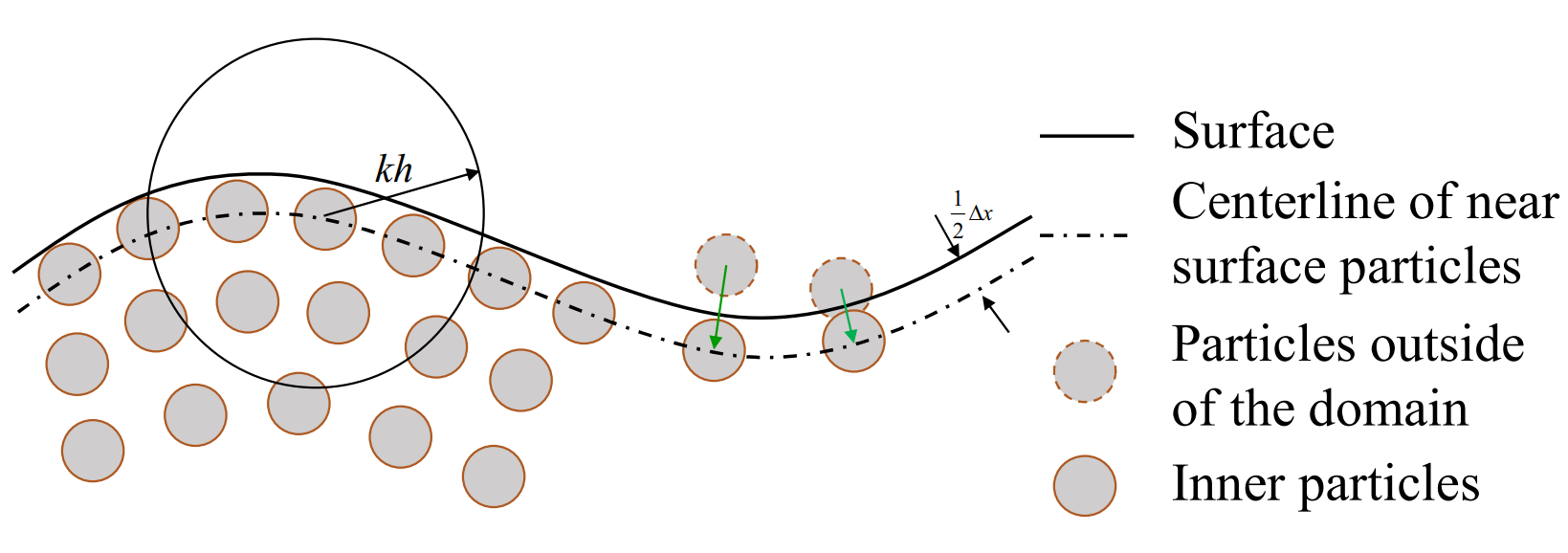}
  \caption{An illustration of surface bounding method \cite{zhu2021cad}.}
  \label{fig: surface-bounding}
\end{figure}

The physics-driven relaxation process can ensure the consistency of the forces between the 
internal particles in order to re-arrange the particles to an isotropic distribution. 
However, since the particles at the boundary of the geometry do not have forces from outside 
to balance the repulsion generated by their internal neighbors, they will escape from the geometry 
boundary during the relaxation process. Thus, Zhu \cite{zhu2021cad} introduced the surface bounding 
method to keep the particles in a body-fitting state during the relaxation process.
The position of boundary particles is reset according to the following equation:
\begin{equation}\label{eq: bounding}
	\mathbf r_i= 
	\begin{cases}
		\mathbf r_i -  \left( \phi_i + \frac{1}{2} \Delta x \right)  \mathbf N_i  \quad  &\phi_i \ge - \frac{1}{2} \Delta x \\
		\mathbf r_i  \quad  \mathrm{otherwise} 
	\end{cases},
\end{equation}
where $\Delta x$ denotes the initial particle spacing, and $\phi_i$ and $\mathbf{N}_i$ are the 
level-set value and normal direction at the position of particle $i$, respectively. As shown in 
Fig.\ref{fig: surface-bounding}, particles near the surface are enforced to locate on the geometry 
surface, thereby achieving a body-fitted distribution.

%%%%%%%%%%%%%%%%%%%%%%%%%%%%%%%%%%%%%%%%%%%%%%%%%%%%%%%%%%%%%
% Section
%%%%%%%%%%%%%%%%%%%%%%%%%%%%%%%%%%%%%%%%%%%%%%%%%%%%%%%%%%%%%
\subsubsection{Static confinement method}
The "Static confinement" method presented by Yu et al. \cite{yu2023level} introduces external forces 
by completing the kernel support of the boundary particles to balance the repulsion of their internal 
neighbors. It could ensure the consistency between the geometry volume and the particle distribution
while achieving the body-fitted particle distribution.

\begin{figure}
  \centering
  \includegraphics[width=4in]{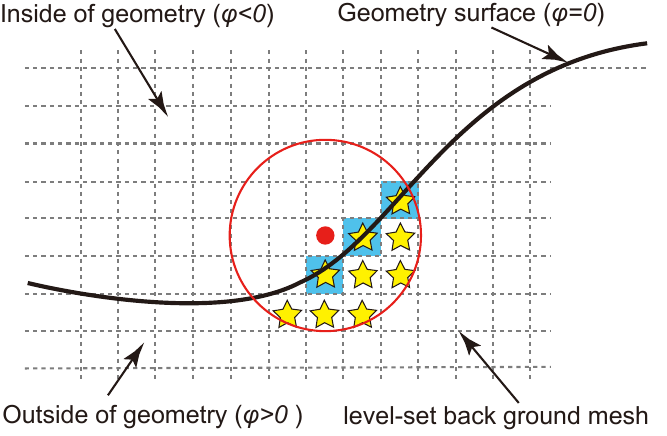}
  \caption{‘Static confinement’ boundary condition for completing kernel support of surface particles. 
  The blue and white cells with yellow stars have partial or full volume contributing to level-set kernel
   support for the target red particle.}
  \label{fig: static-confinement}
\end{figure}

In this method, the kernel support of those boundary particles is provided by their neighbors inside 
the geometry and the background mesh outside the geometry but within the cut-off region, respectively. 
Therefore, the governing equation Eq.\eqref{eq: momentum-sph} in the physics-relaxation process can 
be rewritten as:
\begin{equation}\label{eq: momentum-on-mesh}
    \mathbf{F}_{p, i} 
	= -\frac{2 p_{0} V_i }{m_i}(\sum_{j, \phi_j < 0} \nabla W_{ij} V_j  + \sum_{k, \phi_k > 0} 
 \nabla W_{ik} V_k ).
\end{equation}
where the first term in the bracket of the right-hand-side provides support from the neighbor particles
inside the geometry, while the second term is the support outside of the geometry 
boundary. Here, the $j$ denotes the neighbor particles of particle $i$, and $V_j$ is their volume. 
The $k$ represents the cell centers of the background level-set mesh within the cut-off 
region of particle $i$ but out of geometry, and $V_k$ is the volume of those cells.
The $\phi$ is the level-set value at the particle position or level-set mesh center. 
Meanwhile, the volume fraction contributed to the kernel support by the background mesh needs 
to be calculated according to the following smoothed Heaviside function:
\begin{equation}\label{eq: heaviside function}
H(\phi,\epsilon)= 
\begin{cases}
0  &\phi < -\epsilon \\
\frac{1}{2} + \frac{\phi}{2\epsilon} + \frac{1}{2\pi}\sin(\frac{\pi\phi}{\epsilon})  &-\epsilon<\phi< \epsilon \\
1  &\phi > \epsilon\\
\end{cases},
\end{equation}
Here, $\epsilon$ is $0.75 $ times the resolution of the level-set mesh, which is denoted by $l_f$.
Thus, the second term in Eq.\eqref{eq: momentum-on-mesh} is reformulated as
\begin{equation}\label{eq: kernel gradient from levelset contribution}
    \sum_{k, \phi_k > 0} \nabla W_{ik} V_k = \sum_{k, \phi_k > 0} H(\phi_k,\epsilon) l^{m}_f \nabla_i W_{ik} = I_{i,j},
\end{equation}
Here $l^{m}_f$, $m$ is the dimension of the background mesh, and $l^{m}_f$ is the volume of each cell.
Fig.\ref{fig: static-confinement} illustrates how the background mesh near the geometry boundary contributes
to the kernel support.

By calculating $I_{i,j}$, the summation of kernel gradient value via 
Eq.\eqref{eq: kernel gradient from levelset contribution} 
at each cell center near the boundary while generating the level-set field at the very beginning of
the simulation process,
it is easy to get the missing kernel support of any particle near the boundary by interpolation from 
the predetermined values at cell centers. Thus greatly accelerating the simulation efficiency.

%%%%%%%%%%%%%%%%%%%%%%%%%%%%%%%%%%%%%%%%%%%%%%%%%%%%%%%%%%%%%
% Section
%%%%%%%%%%%%%%%%%%%%%%%%%%%%%%%%%%%%%%%%%%%%%%%%%%%%%%%%%%%%%
\subsection{Boolean operations on geometry based on level-set}
\label{section: boolean operations}

\begin{figure}
  \centering
  \includegraphics[width=3in]{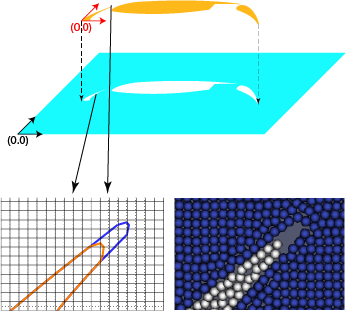}
  \caption{Surface profiles of the airfoil with $\Delta x = L/1000$. The orange line is the geometric 
  boundary of the inner body, while the blue line is that of the outer body. It can be seen that 
  there is a distinct gap at the trailing edge after the two bodies are parsed.}
  \label{fig: 30P30N-levelset}
\end{figure}

As mentioned in the introduction, when importing various geometries of a multi-body system, 
even if the interfaces of the geometries are closely aligned, different parsers will discretize 
the computational domain based on the geometric outer contours, while different outer contours 
lead to different discrete origin points (as shown in Fig.\ref{fig: 30P30N-levelset}, 
the spatial discrete origin point of 
the wing is at the leading edge stagnation point of the flap, 
and the discrete origin point of the fluid domain is at the lower left corner of the computational domain),
which leads to unalignable meshing. 
When the geometry has very small structures that cannot be resolved by the current resolution, 
some geometric information will be discarded based on the resolution. Due to the unalignable meshing, 
the discarded small structures often cannot be the same, thus creating a gap at the multi-body 
interface (as empty space at the trailing edge of the flap in Fig. \ref{fig: 30P30N-levelset}). 
The particle distribution of the multi-body system generated based on such a gapped intersection 
cannot achieve a good quality at the interface. 
Usually, the multi-body systems are composed of solids and fluid domains surrounding them.
Due to the "self-adjustment" mechanism, this gap between fluid and solid particles does not
have a great influence on the simulation results within the Lagrangian framework, as mentioned before.
However, under the framework of Eulerian form, since the initially distributed particles 
cannot adjust their positions by themselves since the simulation starts, 
if the initial particle distribution of the multi-body system contains gaps, 
it will have a greater impact on the accuracy of the simulation result, and even cause the crash.

In order to solve this problem, we perform Boolean operations on different geometries in 
the multi-body system by marking the inclusion relationships of different geometries and their 
corresponding geometric information. In this case, a solid wrapped by a fluid field is usually 
selected as a single geometry, and a corresponding level-set field is obtained as usual. 
The geometric information of the fluid domain is obtained by performing Boolean operations 
on the solid level-set field and the geometric information of the entire computational domain
(as shown in Fig.\ref{fig: 30P30N-bool}). 
The interface between the fluid domain and the solid can be directly represented by the outer 
contour of the solid. Thus, there will be no gap problem in the interface caused by different 
origin points of the mesh discretization. 
The specific geometric characteristics are expressed as follows: 

\begin{figure}
  \centering
  \includegraphics[width=4in]{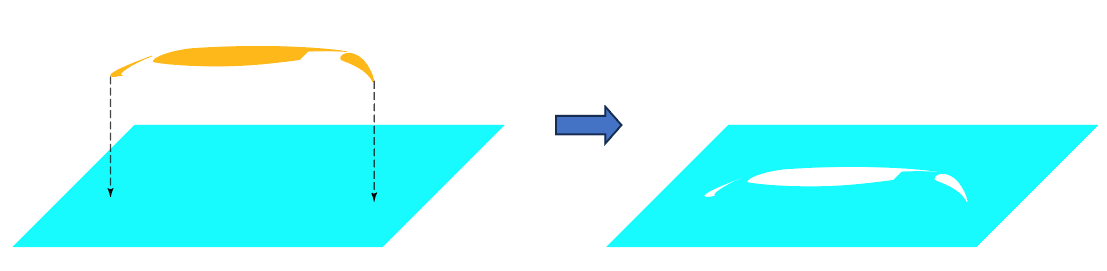}
  \caption{The illustration of Boolean operations on geometry: the orange line in the left picture 
  is the geometric boundary of the inner body, while the blue line is that of the outer boundary 
  of the outer body. Through the subtract operation, the complete geometric boundary of the outer 
  body can be obtained as the right picture.}
  \label{fig: 30P30N-bool}
\end{figure}

The signed distance of fluid domain indicated as blue in  Fig.\ref{fig: 30P30N-bool} for any point A can be expressed as follows:

\begin{equation}\label{eq: signed distance}
D_{A_{\textbf{signed}}}= 
\begin{cases}
-d_A  &A \in \cup \notin \Gamma \\
0 & A = \Gamma \\
\end{cases},
\end{equation}

Here, $d_A$ is the signed distance from any point A to the solid boundary under solid level-set 
field, $\Gamma$ is the out boundary of the solid geometry. While, Other geometric information 
can be obtained based on the signed distance $D_{A_{\textbf{signed}}}$.

%%%%%%%%%%%%%%%%%%%%%%%%%%%%%%%%%%%%%%%%%%%%%%%%%%%%%%%%%%%%%
% Section
%%%%%%%%%%%%%%%%%%%%%%%%%%%%%%%%%%%%%%%%%%%%%%%%%%%%%%%%%%%%%
\subsection{Complex relaxation}\label{section: complex-relaxation}
After obtaining the geometric information of the multi-body system without gaps, 
the next step is generating an isotropic and body-fitting particle distribution for 
the multi-body system, especially how to ensure the quality of particle distribution at the 
interface of multi-body systems. As described in the introduction, many methods fail to produce 
body-fitting particle distributions for complex geometries, such as using ghost 
particles or manually arranging boundary particles. Some methods solve the isotropic and 
body-fitting particle distribution for complex geometries, but the arrangement of boundary 
particles remains a challenge. By using a level-set field to complete the kernel function support 
at the boundary, Yu \cite{yu2023level} has greatly improved the distribution 
quality of near-surface particles. 
Nevertheless, due to its reliance on interpolation approximation, 
this method falls short of attaining a high level of zero-order consistency at the boundaries 
of intricate geometries. Negi \cite{negi2021algorithms} attempted to generate high-quality 
particle distribution for multi-body systems, but the quality of particle distribution 
at the interface is not well guaranteed.

\begin{figure}
  \centering
  \includegraphics[width=4in]{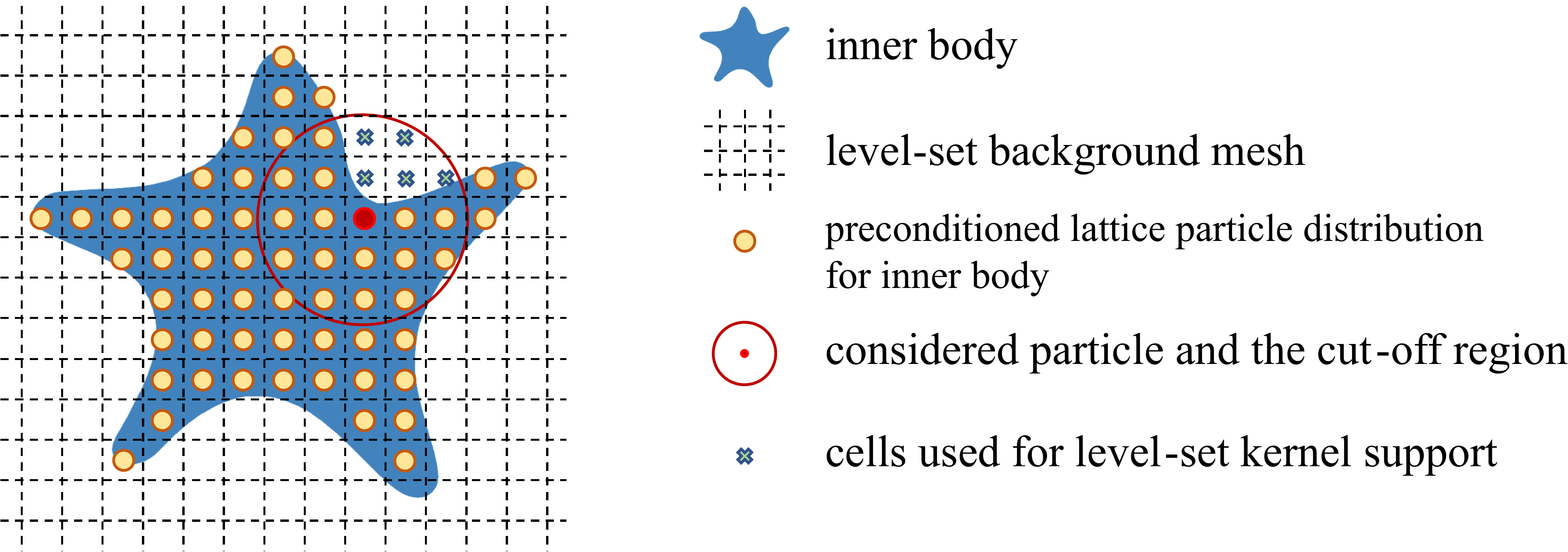}
  \caption{The initial Lattice particle distribution within the inner body. The cut-off region of 
  the considered particle adopts the static confinement method to supplement kernel support.}
  \label{fig: inner-relaxation-illustration}
\end{figure}

\begin{figure}
  \centering
  \includegraphics[width=4in]{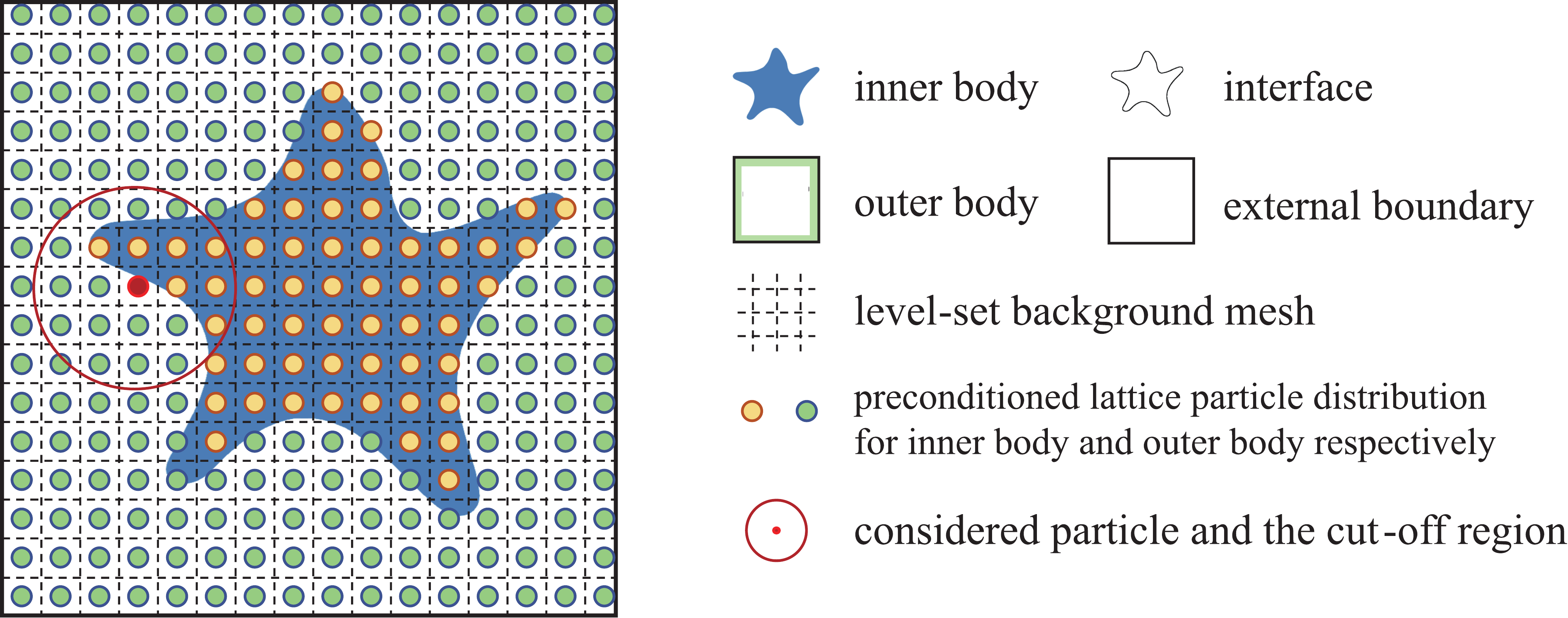}
  \caption{Illustration of complex relaxation for a multi-body system. The considered particle is 
  interpolated by neighbor particles in its own internal configuration and contact bodies.}
  \label{fig: complex-relaxation-illustration}
\end{figure}

Therefore, to enhance the quality of particle distribution at the interface of 
the multi-body systems, we combined some previous work and proposed a multi-body particle 
relaxation method, which is represented as the complex relaxation method in this manuscript. 
In the process of implementing the complex relaxation method, 
it is necessary to perform physics-driven relaxation processes on solid particles 
and fluid particles simultaneously. In each time step, physics-driven relaxation is first 
carried on the internal solid particles, as well as the level-set method used for boundary processing, 
as illustrated in Fig.\ref{fig: inner-relaxation-illustration}
Then, the external fluid particles are performed physics-driven relaxation, 
and its boundary processing is divided into two parts. 
Fluid particles close to the solid boundary need to search for solid 
particles within its support domain for kernel function calculation, while fluid 
particles close to the boundary of the computational domain employed the level-set method 
for kernel function calculation, as depicted in Fig.\ref{fig: complex-relaxation-illustration}.
Therefore, the governing equation for physical relaxation can be rewritten as:

\begin{equation}\label{eq: momentum-complex-relaxation}
    \mathbf{F}_{p, i} 
	= -\frac{2  V_i }{m_i}
 (\sum_{j, \phi_j > 0} p_{0, j} \nabla W_{ij} V_j  + \sum_{k, \phi_k < 0} p_{0, k} \nabla W_{ik} V_k ).
\end{equation}

In this equation, $j$ and $k$ represent the particles of the outer body and the contact inner bodies 
within the cut-off region of target particle $i$ from the outer body.

With this approach, the zero-order consistency of fluid particles at the boundaries 
of multi-body systems is qualitatively improved, which is crucial for particle methods 
in the Eulerian framework.

%%%%%%%%%%%%%%%%%%%%%%%%%%%%%%%%%%%%%%%%%%%%%%%%%%%%%%%%%%%%%
% Section
%%%%%%%%%%%%%%%%%%%%%%%%%%%%%%%%%%%%%%%%%%%%%%%%%%%%%%%%%%%%%
\subsection{Algorithm}

The workflow for a physics-driven complex relaxation process for the multi-body system is presented 
in Algorithm \ref{alg: 1}.

\begin{algorithm}[htb!]
    Setup parameters and initialize the physics-driven relaxation\;
    Predefine or import geometries for the inner solid bodies and parse the polygon mesh\;
    Build the geometry of the outer fluid body using the Boolean operations\;
    Initialize the background Cartesian mesh and the level-set function\;
    Generate a preconditioned Lattice particle distribution for all bodies\;
    \While{simulation termination condition is not satisfied}
        {
        Calculate the pressure force $\mathbf{F}_{p, in}$ with static confinement method for 
        the inner solid bodies according to Eq. \eqref{eq: momentum-on-mesh}\;
        Set the time-step $\Delta t_{in}$ for the inner solid bodies according to 
        Eq. \eqref{eq: time-step}\;
        Update position of inner solid bodies' particles $\mathbf{r}^{n+1}_{in}$ according to 
        Eq.\eqref{eq: update-position}\;
        Apply the surface bounding method to reposition particles outside the inner solid bodies 
        onto the surfaces according to Eq.\eqref{eq: bounding}\;
        Calculate the pressure force $\mathbf{F}_{p, out}$ for the outer fluid body. For the 
        interior particles and the particles near the inner boundary (i.e. the interface with the 
        solid body), the pressure force is calculated by Eq.\eqref{eq: momentum-complex-relaxation}, 
        while the particles located at the external boundary are supplemented by the static 
        confinement method according to Eq. \eqref{eq: momentum-on-mesh}\;
        Set the time-step $\Delta t_{out}$ for the outer fluid body according to 
        Eq.\eqref{eq: time-step}\;
        Update position of outer fluid body's particles $\mathbf{r}^{n+1}_{out}$ according to 
        Eq.\eqref{eq: update-position}\;
        Constrain external boundary particles of the outer fluid body onto the non-contact surface 
        according to Eq.\eqref{eq: bounding}\;
        Update the particle-neighbor list, kernel values, and gradient\;
        Update the particle configuration\;
        }
    Terminate the simulation.
    \caption{Physics-driven complex relaxation for multi-body systems}
    \label{alg: 1}
\end{algorithm}

%%%%%%%%%%%%%%%%%%%%%%%%%%%%%%%%%%%%%%%%%%%%%%%%%%%%%%%%%%%%%
%
% Section
%
%%%%%%%%%%%%%%%%%%%%%%%%%%%%%%%%%%%%%%%%%%%%%%%%%%%%%%%%%%%%%
\section{Numerical examples} \label{section: numerical-examples}

This section is structured into two main categories: particle generation cases and 
physical simulation cases. In the particle generation cases, we conduct a detailed 
comparison of the quality of particle distributions after the generation process, 
contrasting the complex relaxation method introduced in this study with 
the separate relaxation approach, as well as the particle packing algorithm reported 
in the research of Negi et al. \cite{negi2021algorithms}. 
In addition, the physical simulation case shows the numerical simulation results of the flow around an 
airfoil, showcasing the superiority of the complex relaxation method presented in this paper. 
For all cases presented, we employ the 5th-order Wendland C2 kernel function, with a smoothing 
length $h_f = 1.3 \Delta x$ for fluids and $h_s = 1.05 \Delta x$ for solids. In the following cases, 
the separate relaxation refers to undergoing independent particle generation 
processes for each body. Unless otherwise specified, Boolean operations on geometry
introduced in Section \ref{section: boolean operations} are a prerequisite for parsing multiple bodies. 
All simulations in this work are carried out on an Intel Core(TM) CPU i9-10900 
2.80GHZ Desktop computer with 32GB RAM and Windows 10 system.
%%%%%%%%%%%%%%%%%%%%%%%%%%%%%%%%%%%%%%%%%%%%%%%%%%%%%%%%%%%%%
% Section
%%%%%%%%%%%%%%%%%%%%%%%%%%%%%%%%%%%%%%%%%%%%%%%%%%%%%%%%%%%%%
\subsection{Relaxation results} \label{section: relaxation-cases}
%%%%%%%%%%%%%%%%%%%%%%%%%%%%%%%%%%%%%%%%%%%%%%%%%%%%%%%%%%%%%
% Section
%%%%%%%%%%%%%%%%%%%%%%%%%%%%%%%%%%%%%%%%%%%%%%%%%%%%%%%%%%%%%
\subsubsection{2D airfoil 30P30N} \label{section: 30P30N}

\begin{figure}
  \centering
  \includegraphics[width=4in]{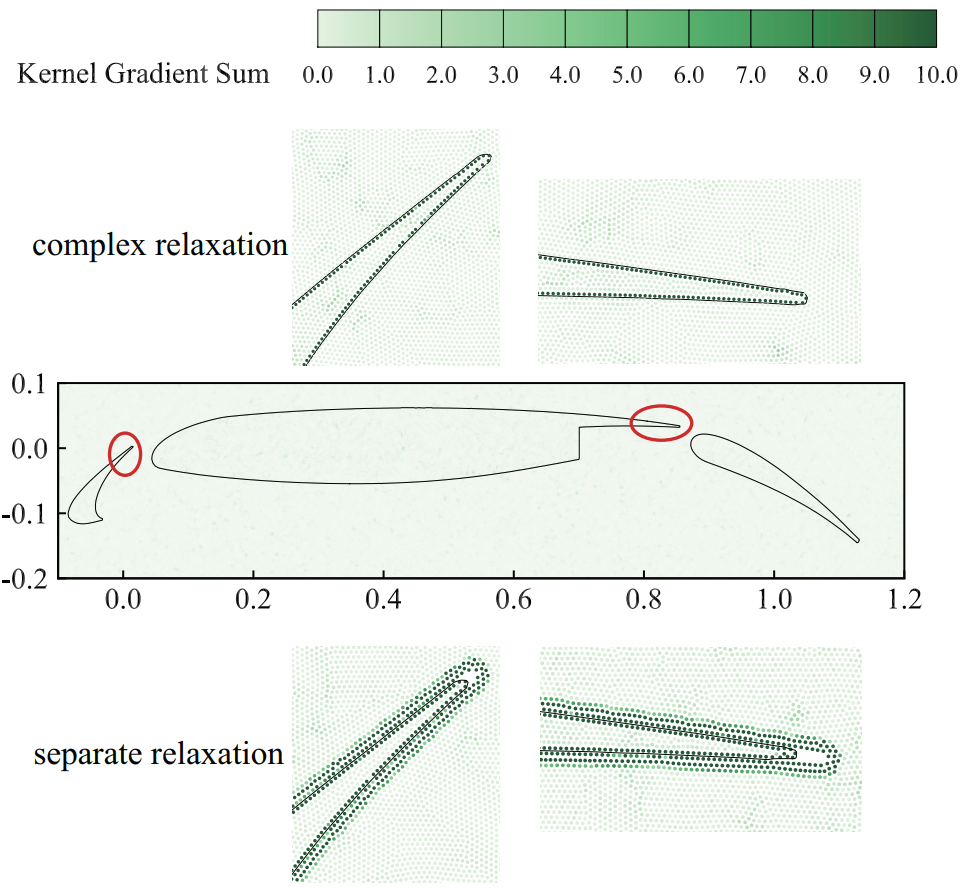}
  \caption{Comparison of kernel gradient summation for the 30P30N airfoil between complex 
  relaxation and separate relaxation with $\Delta x = L/1000$.}
  \label{fig: airfoil-consistency}
\end{figure}

Fig.\ref{fig: airfoil-consistency} presents the particle distribution for the 2D airfoil 30P30N and its 
surrounding water body. Here, we also carried out another test case: generating particles separately 
in the solid airfoil and the fluid body without Boolean operations, to further demonstrate the 
consequences of particle distributions in the case of non-coincidence of level-sets between the 
inner solid and outer fluid bodies mentioned in \ref{section: boolean operations}.

In this scenario, the solid airfoil body is treated only as a boundary to the 
fluid and does not actively participate in the physical simulation. Therefore, when evaluating zero-order 
consistency for the inner solid body, only its internal solid particles are considered. However, 
for the outer fluid body, the information of near-surface particles includes contributions not only from 
their internal fluid particles but also from the contact solid particles, i.e. the wall (airfoil) particles.
Zero-order consistency implies that the kernel function satisfies $\int \nabla W(x-x', h)\mathrm{d}x' = 0$, 
where $x$ is the position vector. In SPH formulation, this can be discretized into the kernel gradient 
summation (KGS), expressed as:
\begin{equation}\label{eq: kernel gradient sum}
KGS_i= 
\begin{cases}
\sum_{j, \phi_j < 0} \nabla_i W_{ij} V_j  &{i \in \textnormal{inner solid}}\\
\sum_{j, \phi_j > 0} \nabla_i W_{ij} V_j + \sum_{k, \phi_k < 0} \nabla_i W_{ik} V_k  
&{i \in \textnormal{outer fluid}}\\
\end{cases},
\end{equation}
In this equation, the first case applies when particle $i$ is from the inner solid body, and neighboring 
particle $j$ is from the same body. The second case applies when particle $i$ is from the outer fluid body, 
with particle $j$ belonging to its internal fluid particles and particle $k$ belonging to the contact 
solid particles.

As shown in Fig.\ref{fig: airfoil-consistency}, both the complex and separate relaxation methods can 
maintain a uniform distribution for the inner solid body (comprising the airfoil itself). This is because 
the particle generation processes for the solid are identical, and the previous boundary treatments using 
the surface bounding method combined with the static confinement method are sufficient for a single body. 
However, for the outer water body, separate relaxation with Boolean operations on geometry, 
while effective in eliminating void regions near sharp structures as discussed in 
\ref{section: boolean operations}, still struggles to homogenize particle locations at the interface. 
This is evidenced by the high kernel gradient summation values. In contrast, the complex relaxation 
method effectively addresses this issue, leading to a more homogeneous particle distribution globally, 
even in the sharp trailing edge regions.

\begin{figure}
  \centering
  \includegraphics[width=3in]{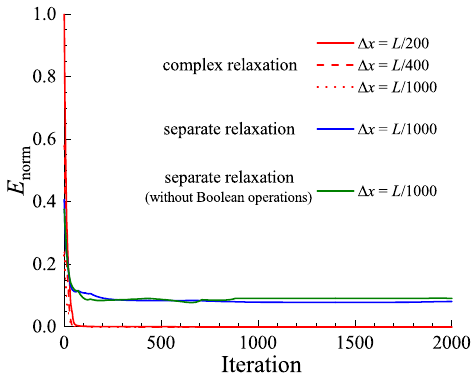}
  \caption{Normalized kinetic energy comparison between the complex and separate relaxation processes 
  for the 30P30N airfoil.}
  \label{fig: airfoil-kinetic-energy}
\end{figure}

Kinetic energy serves as an indicator of the convergence of the particle generation process \cite{zhu2021cad}.
Under the ideal zero-order consistency, the particle distribution should reach an equilibrium state 
where the interaction forces between particles are balanced, resulting in smaller particle velocities. 
Conversely, if the particles are unevenly distributed, the imbalanced force will cause higher particle 
velocities. Therefore, the kinetic energy of uniform and consistent particle distribution tends to 
stabilize and converge to a lower value. It is defined by the equation
\begin{equation}\label{eq: kinetic energy}
    E = \frac{1}{2} \sum_i{m_iv_i^2},
\end{equation}
where $v_i = \frac{\mathrm{d}\mathbf{v_i}}{\mathrm{d}t} \cdot \Delta t$ and $m_i$ denoting the velocity 
and mass of the particle $i$, respectively. Given the differences between the relaxation methods 
are primarily at the outer body interface, to emphasize the contrast in kinetic energy values after 
normalization, only the first-layer fluid particles near the wall are considered as target particles 
to calculate the summation of kinetic energy.
Fig.\ref{fig: airfoil-kinetic-energy} displays the comparison of kinetic energy between the complex 
and separate relaxation processes for the 30P30N airfoil, normalized by the largest kinetic energy 
value at the initial time, with $L$ equal to 1. Notably, the kinetic energy of complex relaxation 
across three resolutions all converge towards zero, whereas those for separate relaxation remain 
significantly higher.

%%%%%%%%%%%%%%%%%%%%%%%%%%%%%%%%%%%%%%%%%%%%%%%%%%%%%%%%%%%%%
% Section
%%%%%%%%%%%%%%%%%%%%%%%%%%%%%%%%%%%%%%%%%%%%%%%%%%%%%%%%%%%%%
\subsubsection{2D starfish} \label{section: 2D-starfish}

\begin{figure}
	\centering
	\begin{subfigure}[b]{0.49\textwidth}
		\centering
		\includegraphics[width=\textwidth]{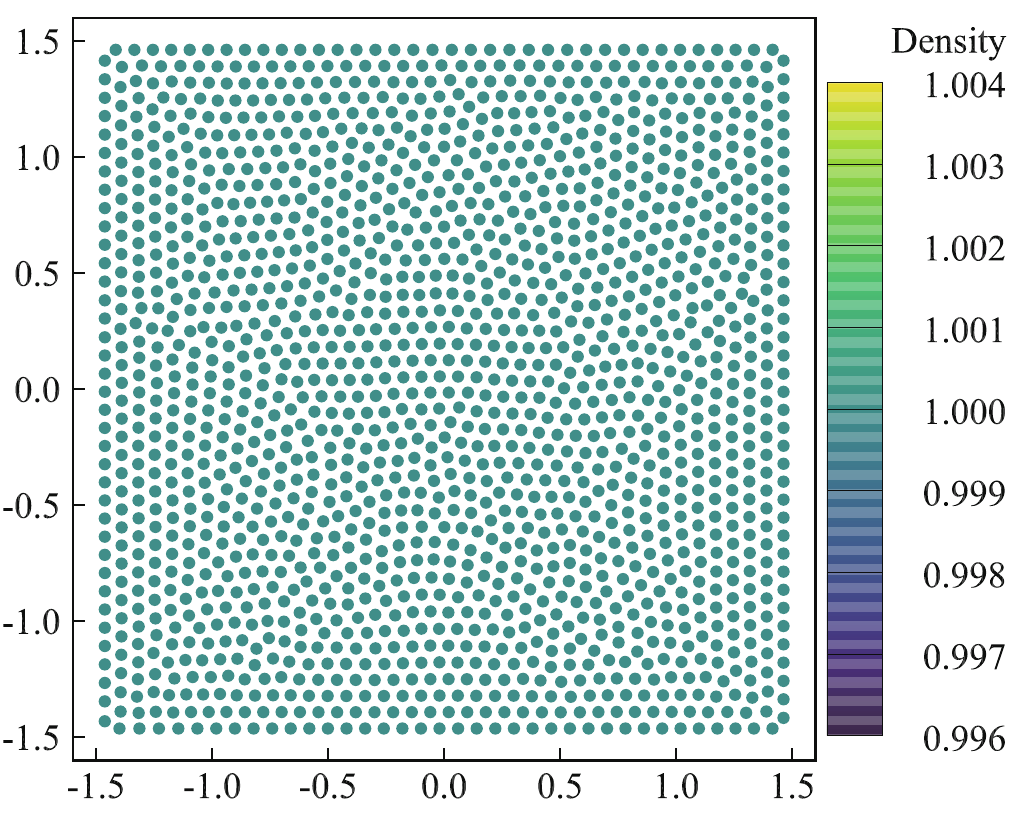}
		\caption{}
		\label{fig: starfish-complex-density}
	\end{subfigure}
	%\newline 
	\begin{subfigure}[b]{0.49\textwidth}
		\centering
		\includegraphics[width=\textwidth]{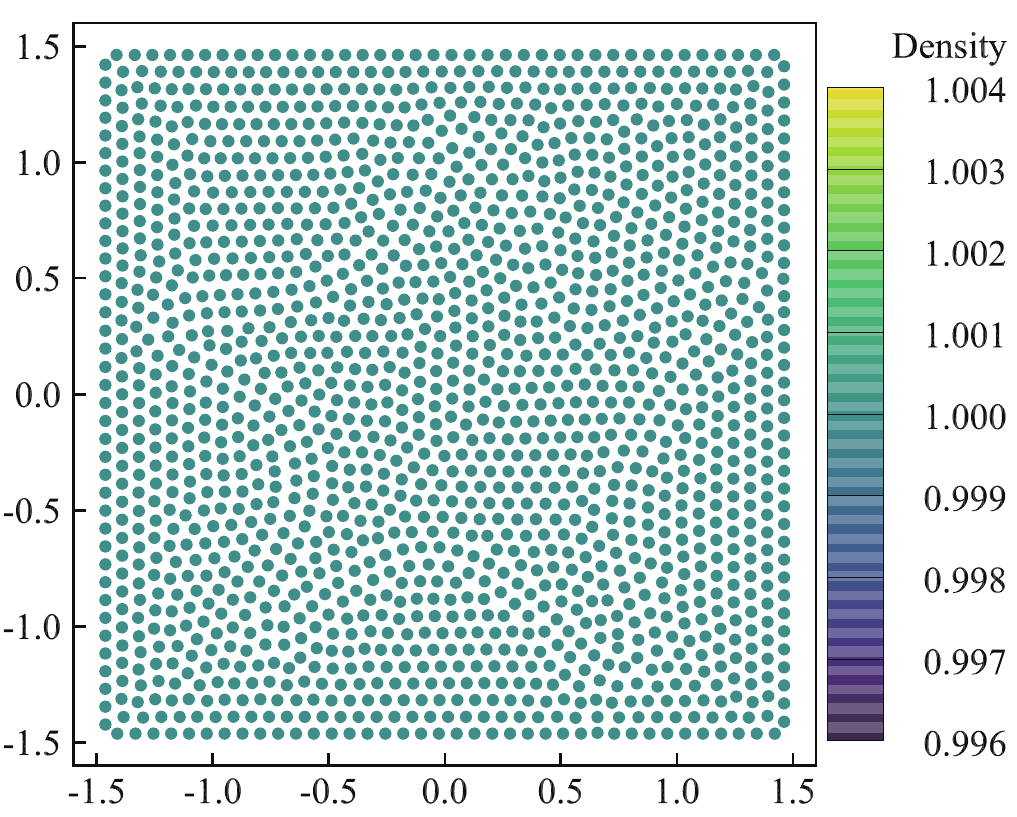}
		\caption{}
		\label{fig: starfish-separate-density}
	\end{subfigure}
    %\newline 
	\begin{subfigure}[b]{0.5\textwidth}
		\centering
		\includegraphics[width=\textwidth]{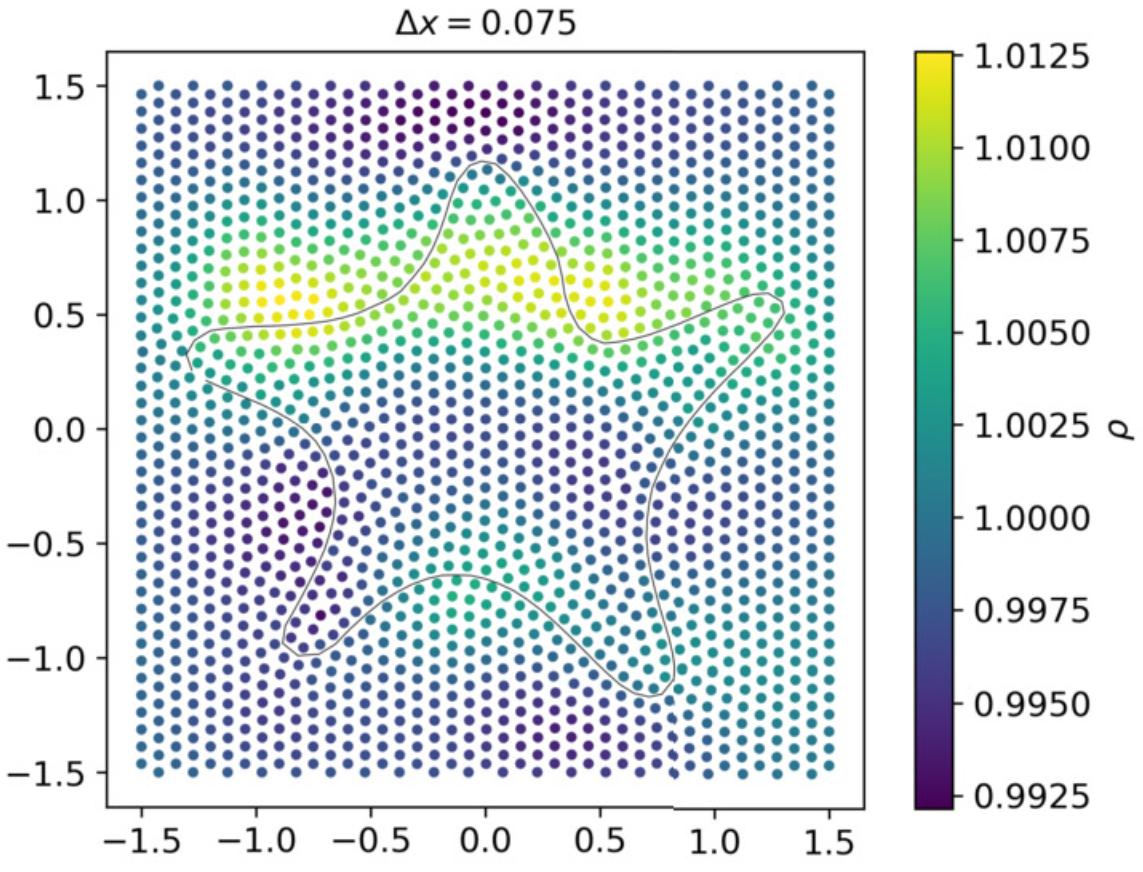}
		\caption{}
		\label{fig: starfish-negi-density}
	\end{subfigure}
	\caption{Density distribution for the starfish with $\Delta x = L/20$. (a) and (b) are obtained 
 by complex relaxation and separate relaxation, respectively. (c) is given in 
 Ref.\cite{negi2021algorithms}.}
	\label{fig: starfish-density}
\end{figure}

\begin{figure}
	\centering
	\begin{subfigure}[b]{0.49\textwidth}
		\centering
		\includegraphics[width=\textwidth]{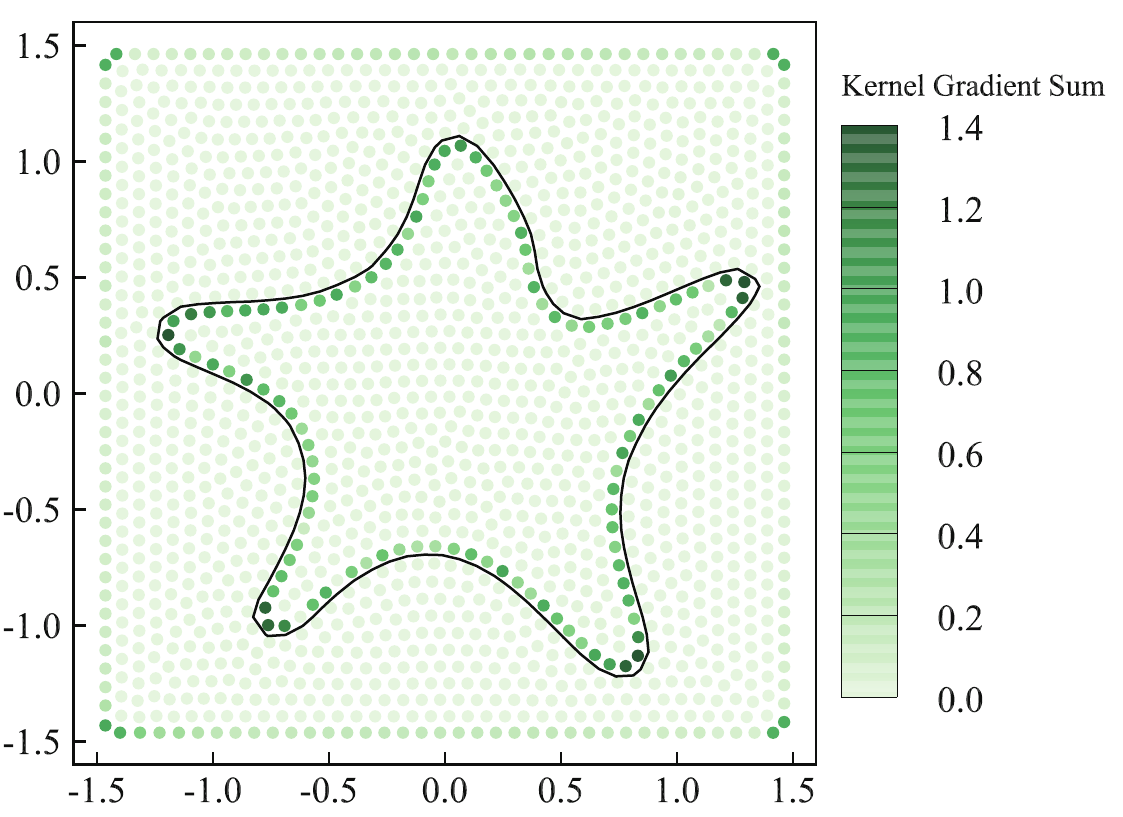}
		\caption{}
		\label{fig: starfish-complex-consistency}
	\end{subfigure}
	%\newline 
	\begin{subfigure}[b]{0.49\textwidth}
		\centering
		\includegraphics[width=\textwidth]{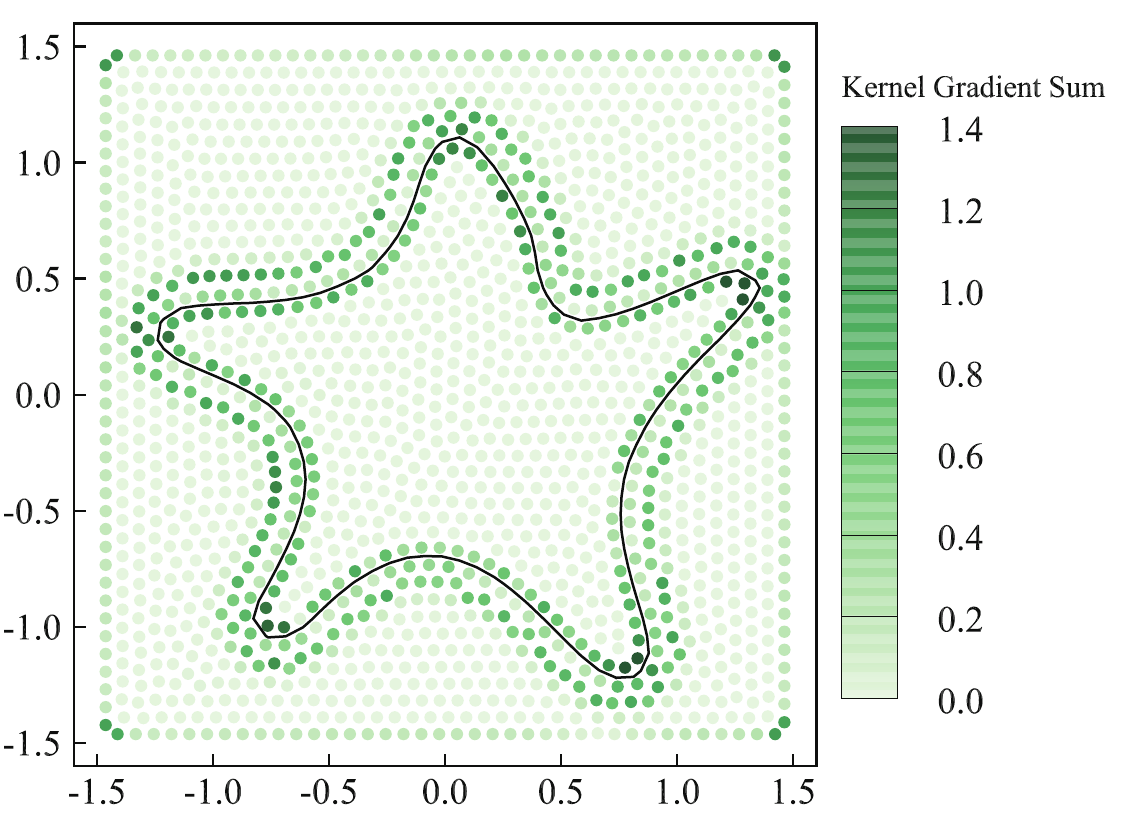}
		\caption{}
		\label{fig: starfish-separate-consistency}
	\end{subfigure}
	\caption{Kernel gradient summation for the starfish with $\Delta x = L/20$ by (a) complex relaxation and (b) separate relaxation.}
	\label{fig: starfish-consistency}
\end{figure}

In the 2D starfish case study, a particle spacing $\Delta x = 0.075$ ($\Delta x = L/20$ and $L$ is 
the side length of the fluid domain, equal to 1.5 here) is initially set to illustrate the density 
distribution in a two-body system resulting from both complex and separate relaxation methods, 
as well as the result reported in Ref.\cite{negi2021algorithms}, as shown 
in Fig.\ref{fig: starfish-density}. The comparison indicates that irrespective of whether the particle 
relaxation process is conducted jointly or separately in multiple bodies, the density distributions 
achieved via the physics-driven relaxation method with constant background pressure as well as the 
application of boundary modification are more homogeneous than that presented in the reference. When 
compared to the separate relaxation result in Fig.\ref{fig: starfish-separate-density}, it is evident 
that complex relaxation in Fig.\ref{fig: starfish-complex-density} achieves a more uniform particle 
distribution at equivalent resolutions, particularly at the interface between the two bodies. 
In addition, the kernel gradient summation obtained by complex relaxation in 
Fig.\ref{fig: starfish-complex-consistency} of the outer fluid body at the interface is smaller than that 
obtained by separate relaxation, which brings additional benefits for further physical simulation. 
It is indicated that the fluid particles at the interface are relaxed through a complex relaxation 
procedure that integrates information from the particles of the inner solid body. 
This process enhances the zero-order consistency in this region.

\begin{figure}
	\centering
	\begin{subfigure}[b]{0.49\textwidth}
		\centering
		\includegraphics[width=\textwidth]{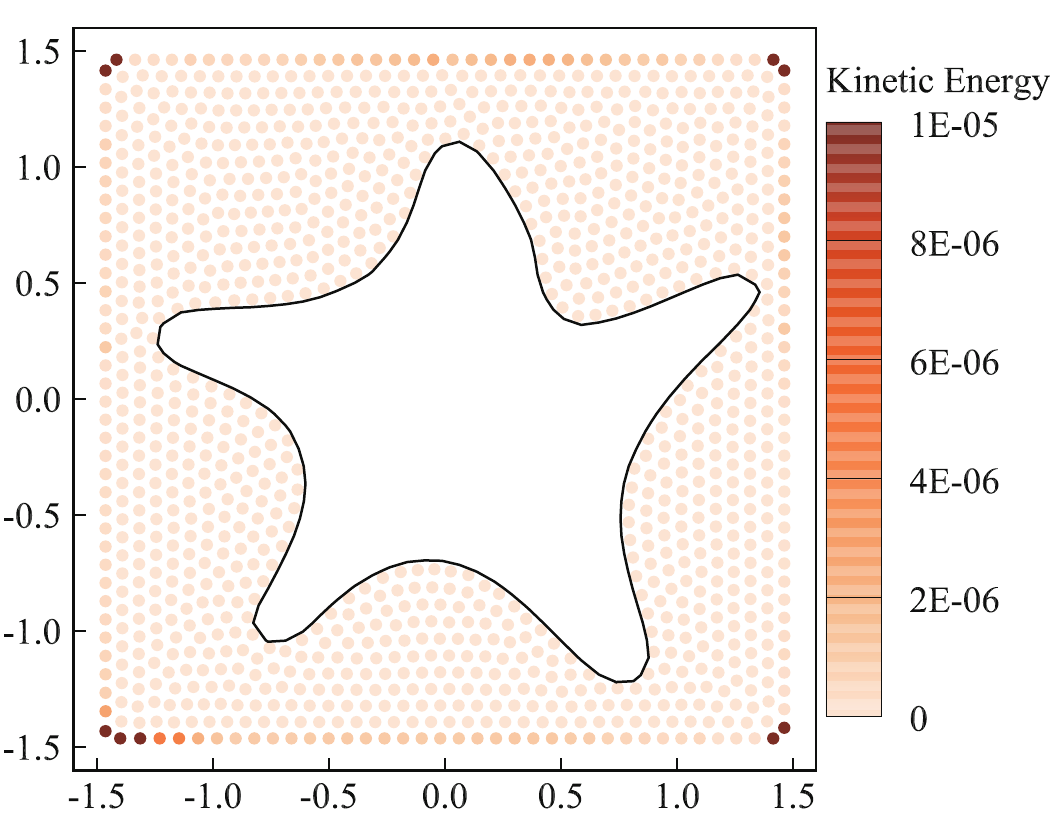}
		\caption{}
		\label{fig: starfish-complex-energy}
	\end{subfigure}
	%\newline 
	\begin{subfigure}[b]{0.49\textwidth}
		\centering
		\includegraphics[width=\textwidth]{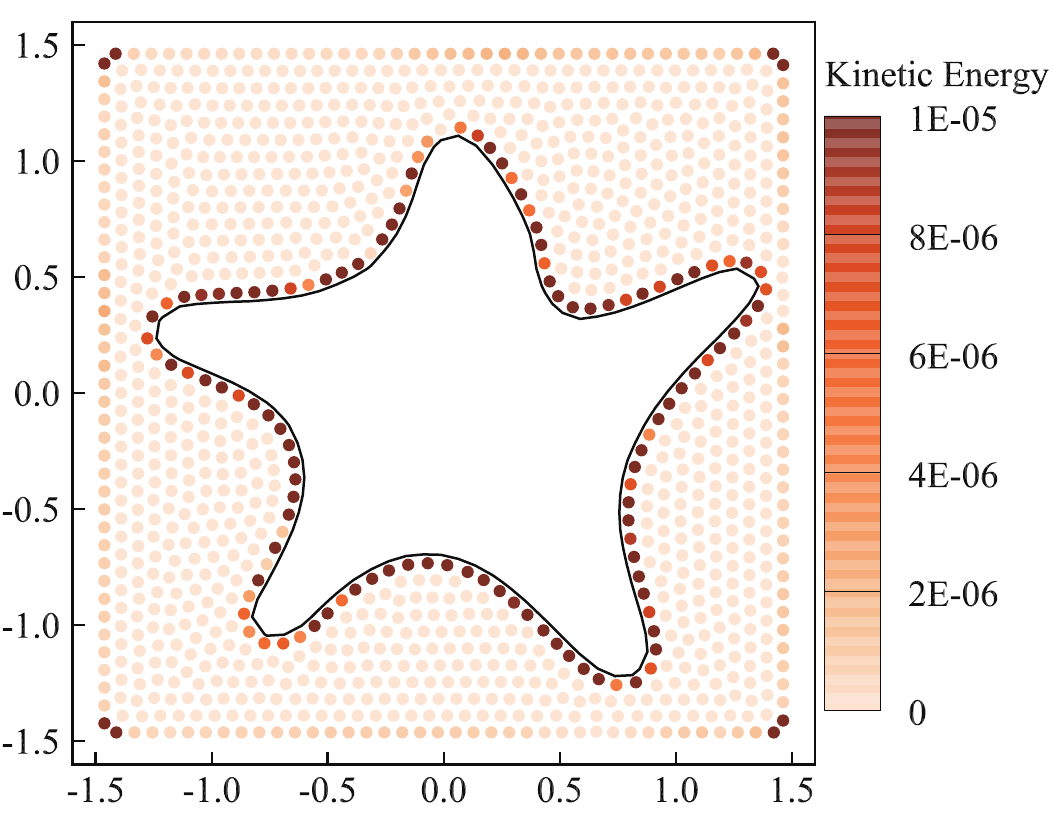}
		\caption{}
		\label{fig: starfish-separate-energy}
	\end{subfigure}
	\caption{Kinetic energy distribution for the starfish with $\Delta x = L/20$ by (a) complex relaxation and (b) separate relaxation.}
	\label{fig: starfish-energy}
\end{figure}

\begin{figure}
  \centering
  \includegraphics[width=3in]{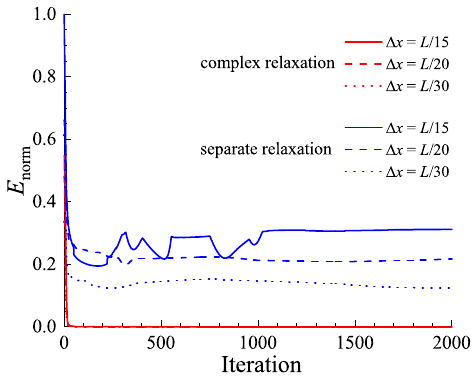}
  \caption{Normalized particle kinetic energy during the physics-driven relaxation process at 
 various resolutions for the starfish.}
  \label{fig: starfish-kinetic-energy}
\end{figure}

To further validate the assertion that the significant difference in particle kinetic 
energy between the two relaxation methods is primarily located at the interface of the 
fluid body, Fig.\ref{fig: starfish-energy} compares the kinetic energy distribution of 
fluid particles between complex relaxation and separate relaxation. Evidently, the 
fluid particle distribution obtained via the separate relaxation approach exhibits 
distinct high-value regions near the wall, indicating uneven particle distribution. 
Consequently, we focus on the interface layer of fluid particles. As illustrated in 
Fig.\ref{fig: starfish-kinetic-energy}, the normalized kinetic energy of the complex 
relaxation process converges significantly to lower values compared to that of the 
separate relaxation process.

%%%%%%%%%%%%%%%%%%%%%%%%%%%%%%%%%%%%%%%%%%%%%%%%%%%%%%%%%%%%%
% Section
%%%%%%%%%%%%%%%%%%%%%%%%%%%%%%%%%%%%%%%%%%%%%%%%%%%%%%%%%%%%%
\subsubsection{2D zig-zag wall}

\begin{figure}
	\centering
	\begin{subfigure}[b]{0.4\textwidth}
		\centering
		\includegraphics[width=\textwidth]{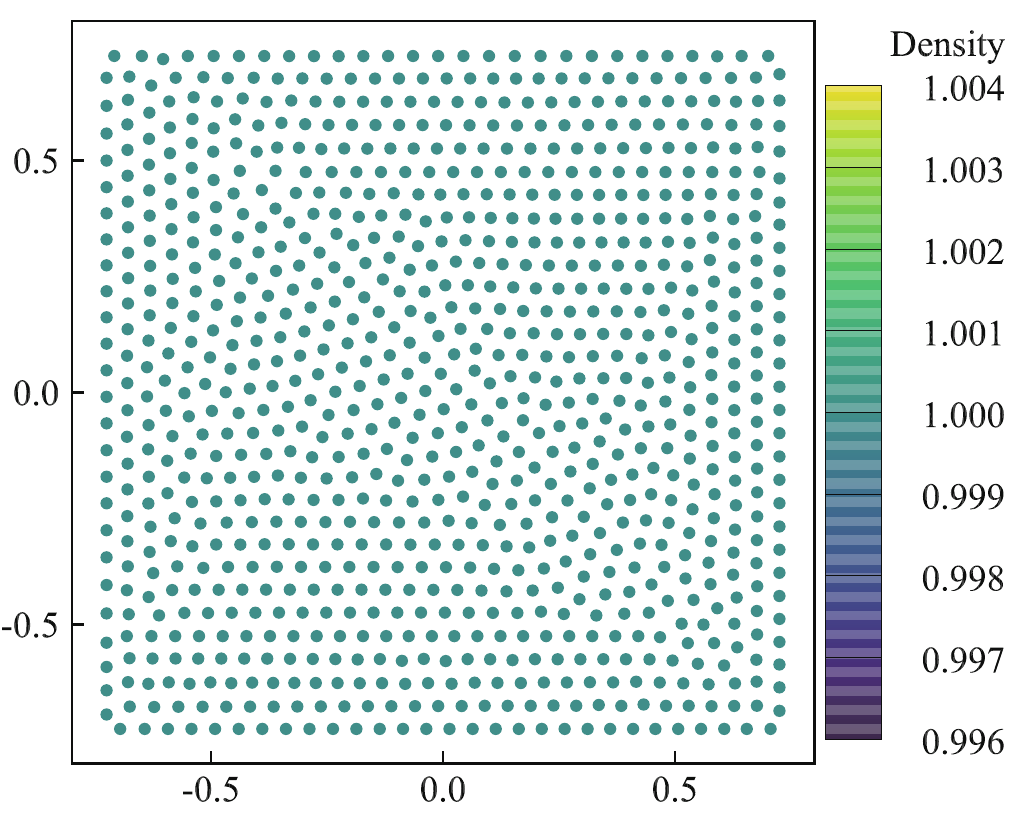}
		\caption{}
		\label{fig: zigzag-complex-density}
	\end{subfigure}
	%\newline 
	\begin{subfigure}[b]{0.4\textwidth}
		\centering
		\includegraphics[width=\textwidth]{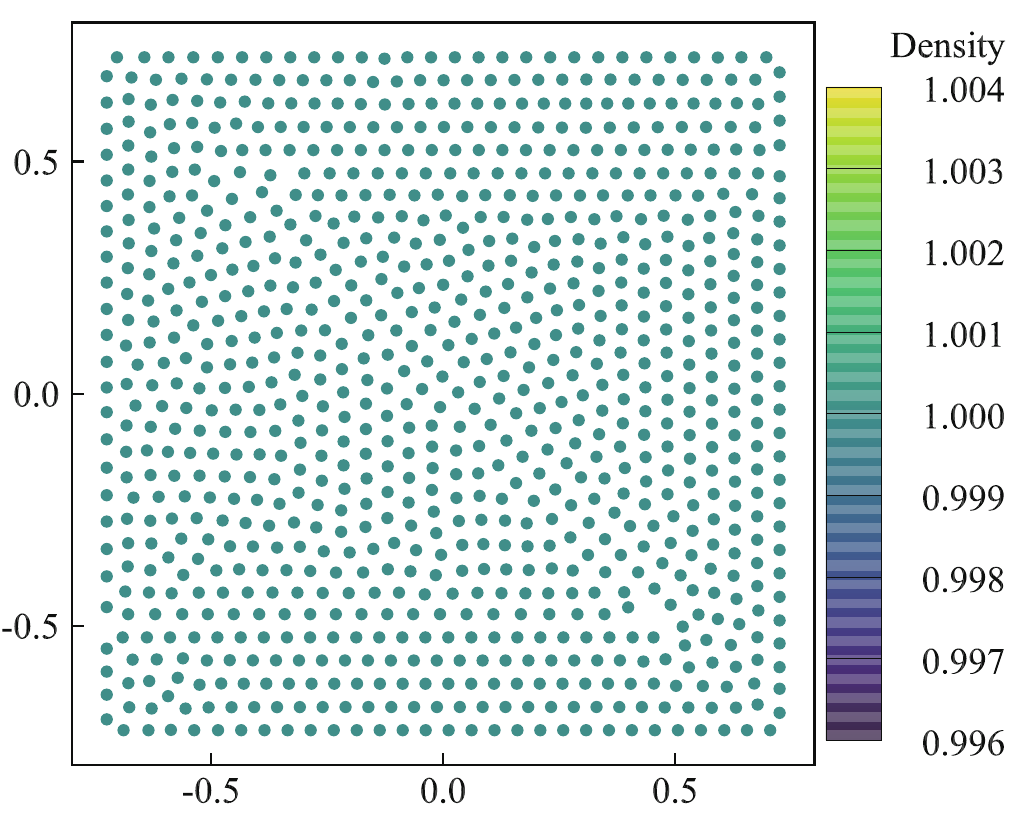}
		\caption{}
		\label{fig: zigzag-separate-density}
	\end{subfigure}
    %\newline 
	\begin{subfigure}[b]{0.5\textwidth}
		\centering
		\includegraphics[width=\textwidth]{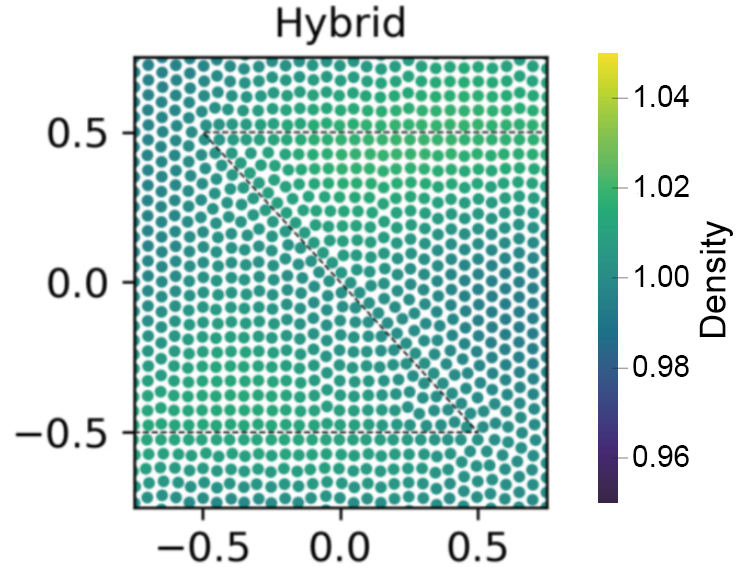}
		\caption{}
		\label{fig: zigzag-negi-density}
	\end{subfigure}
	\caption{Density distribution for the zigzag wall with $\Delta x = L/15$. (a) and (b) are 
 obtained by complex relaxation and separate relaxation, respectively. (c) is given in 
 Ref.\cite{negi2021algorithms}.}
	\label{fig: zigzag-density}
\end{figure}

In the 2D zig-zag wall test case, the comparison of density distribution using different algorithms is 
depicted in Fig.\ref{fig: zigzag-density} with the particle spacing $\Delta x = 0.05$ 
($\Delta x = L/15$ and and $L$ is the side length of the fluid domain, equal to 0.75 here). 
In Fig.\ref{fig: zigzag-negi-density}, derived from the methodology in 
Ref.\cite{negi2021algorithms}, there is a dense concentration of particles at the interface of two 
halves of the zig-zag wall. In contrast, in the particle distribution obtained by the complex 
relaxation method proposed in this paper shown in Fig.\ref{fig: zigzag-complex-density}, the interface 
and internal distribution are almost indistinguishable, which exhibits excellent isotropic 
characteristics. On the other hand, the result from the separate relaxation method, as seen in 
Figure \ref{fig: zigzag-separate-density} reveals two distinct void areas near the sharp corners 
at the interface. This indicates the inability of the two bodies to achieve a uniform distribution 
through independent relaxation processes.

\begin{figure}
	\centering
	\begin{subfigure}[b]{0.49\textwidth}
		\centering
		\includegraphics[width=\textwidth]{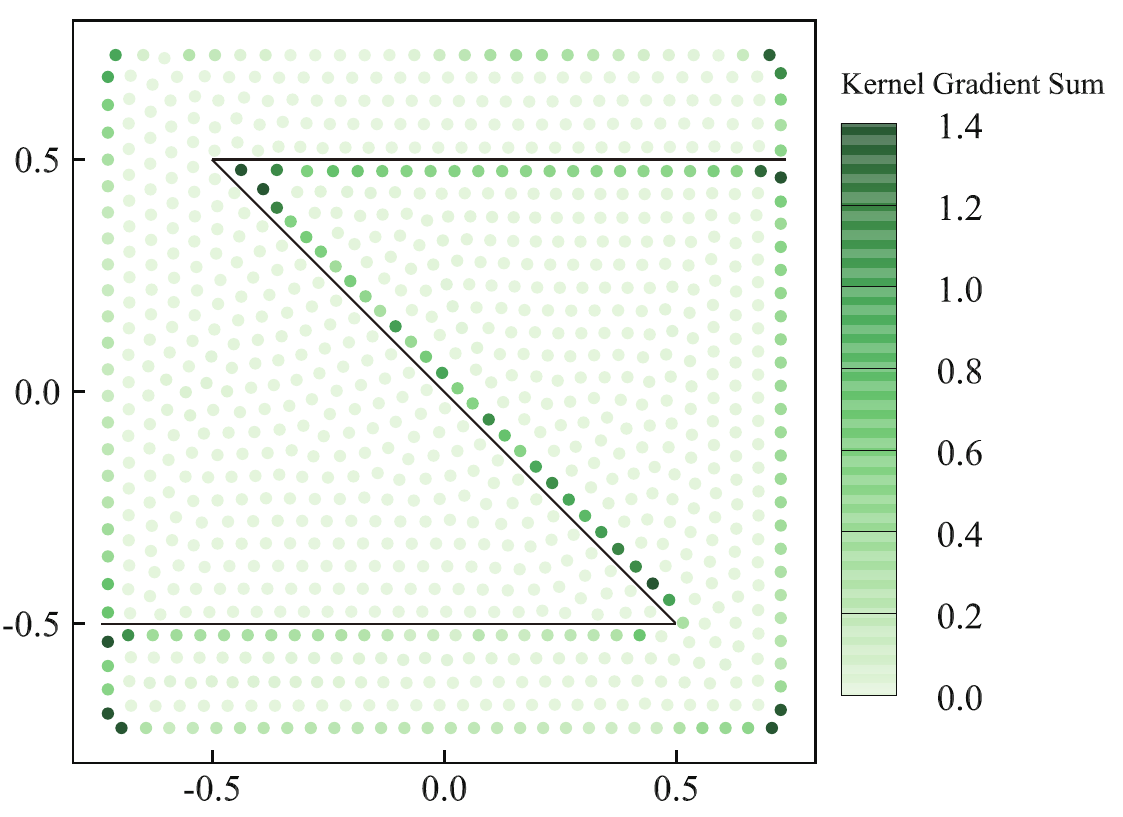}
		\caption{}
		\label{fig: zigzag-complex-consistency}
	\end{subfigure}
	%\newline 
	\begin{subfigure}[b]{0.49\textwidth}
		\centering
		\includegraphics[width=\textwidth]{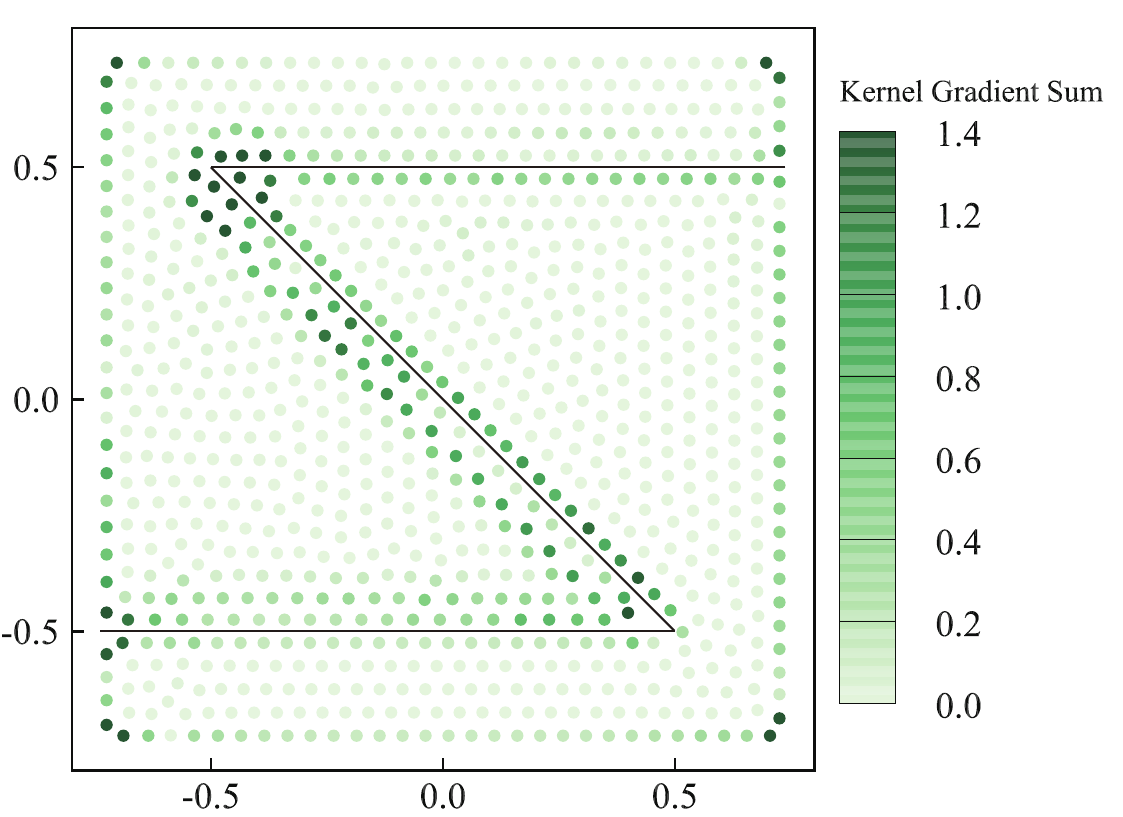}
		\caption{}
		\label{fig: zigzag-separate-consistency}
	\end{subfigure}
	\caption{Kernel gradient summation for the zigzag wall with $\Delta x = L/15$ by (a) complex 
 relaxation and (b) separate relaxation.}
	\label{fig: zigzag-consistency}
\end{figure}

\begin{figure}
  \centering
  \includegraphics[width=3in]{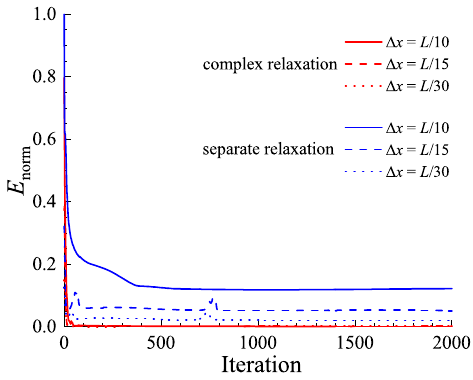}
  \caption{Normalized particle kinetic energy during the physics-driven relaxation process at 
 various resolutions for the zig-zag wall.}
  \label{fig: zigzag-kinetic-energy}
\end{figure}

The kernel gradient summation from the separate relaxation process exhibits a greater deviation from 
zero-order consistency compared to that obtained through complex relaxation, as depicted in 
Fig.\ref{fig: zigzag-consistency}. Besides, the kinetic energy curves show a declining tendency 
with the resolution increasing, as shown in Fig.\ref{fig: zigzag-kinetic-energy}, whereas the 
convergence values of complex relaxation are significantly lower than those of the separate relaxation.

%%%%%%%%%%%%%%%%%%%%%%%%%%%%%%%%%%%%%%%%%%%%%%%%%%%%%%%%%%%%%
% Section
%%%%%%%%%%%%%%%%%%%%%%%%%%%%%%%%%%%%%%%%%%%%%%%%%%%%%%%%%%%%%
\subsubsection{3D Stanford bunny}

\begin{figure}
  \centering
  \includegraphics[width=3in]{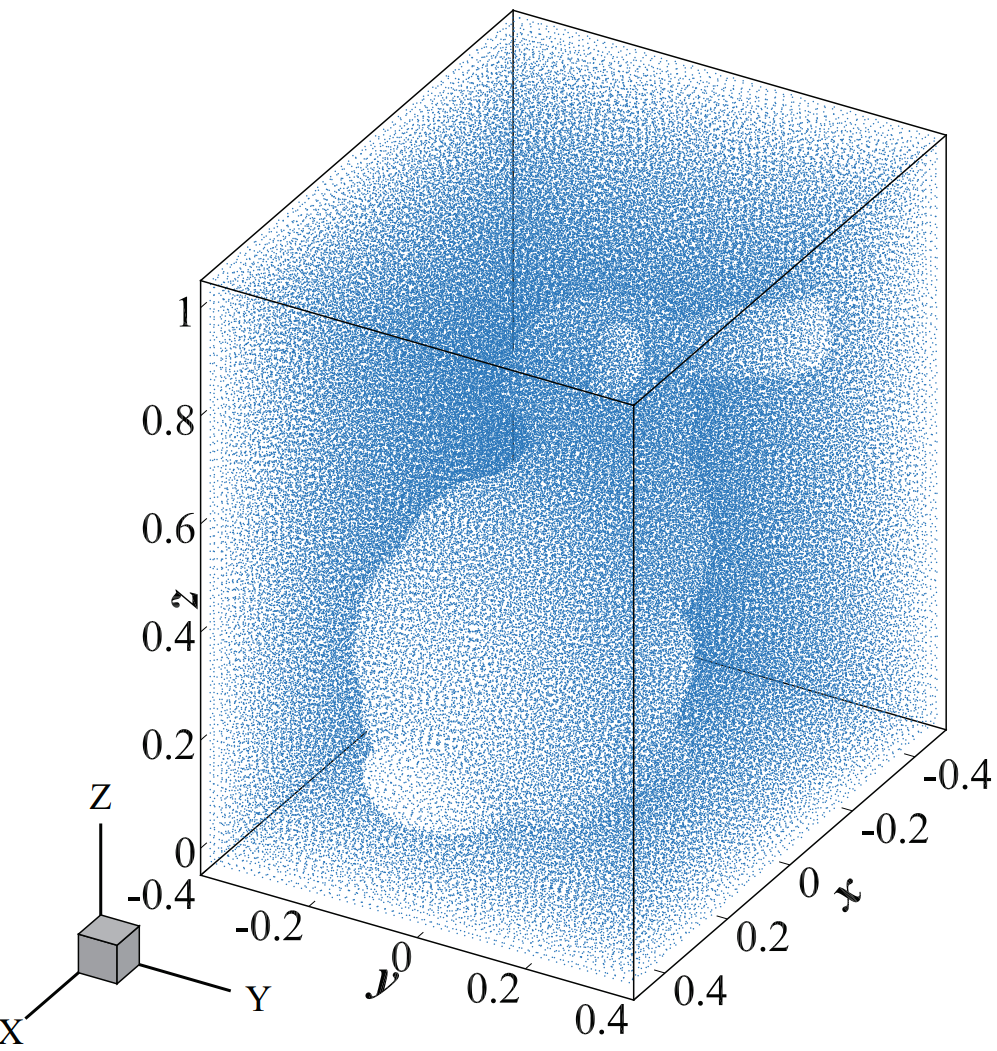}
  \caption{Computational domain of fluid and solid bodies in bunny case.}
  \label{fig: bunny-case-domain}
\end{figure}

\begin{figure}
	\centering
	\begin{subfigure}[b]{0.6\textwidth}
		\centering
		\includegraphics[width=\textwidth]{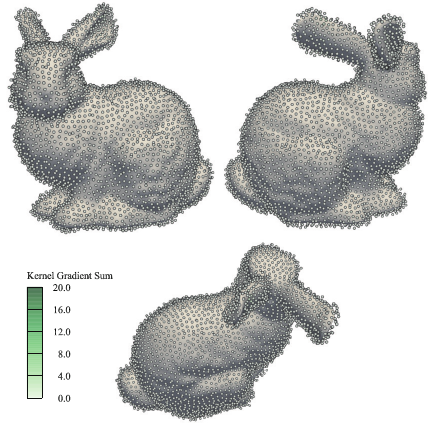}
		\caption{}
		\label{fig: bunny-complex-consistency}
	\end{subfigure}
	%\newline 
	\begin{subfigure}[b]{0.6\textwidth}
		\centering
		\includegraphics[width=\textwidth]{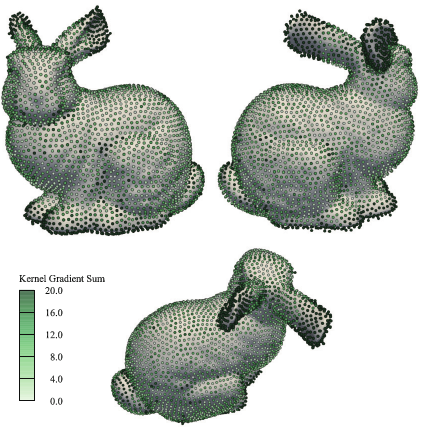}
		\caption{}
		\label{fig: bunny-separate-consistency}
	\end{subfigure}
	\caption{Kernel gradient summation of fluid particles at the outer surface of the bunny 
	with $\Delta x = 0.02$ by (a) complex relaxation and (b) separate relaxation.}
	\label{fig: bunny-consistency}
\end{figure}

This subsection presents the application of the complex relaxation approach introduced in this 
study to a three-dimensional case, involving the geometry of the Stanford bunny 
as the inner body. The computational domain is depicted in Fig.\ref{fig: bunny-case-domain}
and the initial particle spacing of fluid and solid bodies is 0.02. 
Fig.\ref{fig: bunny-consistency} shows the kernel gradient summation of particles at the outer surface 
of the bunny, i.e. the first-layer fluid particles near the interface. The first layer of 
fluid particles obtained through complex relaxation achieves a good zero-order consistency at the 
fluid-solid interface. This case affirms the robustness of the proposed method in ensuring zero-order 
consistency at the interface of the multi-body systems when handling complex three-dimensional geometries.

%%%%%%%%%%%%%%%%%%%%%%%%%%%%%%%%%%%%%%%%%%%%%%%%%%%%%%%%%%%%%
% Section
%%%%%%%%%%%%%%%%%%%%%%%%%%%%%%%%%%%%%%%%%%%%%%%%%%%%%%%%%%%%%
\subsection{Physical simulation results} \label{section: physical-cases}

To demonstrate the superiority of the complex relaxation method presented in this paper over the 
separate relaxation method, we examine the flow around the NACA 5515 airfoil as a test case and conduct 
a physical simulation within the Eulerian framework. The airfoil with a narrow and elongated trailing 
edge shows challenges in particle arrangement. Particles are generally pre-arranged at the geometric 
boundary of the airfoil as seen in previous studies \cite{huang2019kernel, shadloo2011improved}, while 
the particle relaxation method eliminates the need for this complicated step. Upon completion of the 
relaxation process, both solid and fluid particles remain stationary under the Eulerian framework. 
Consequently, the initial particle distribution plays a crucial role in the overall accuracy of the 
simulation results. The detailed theories of the Eulerian SPH can be found in 
Ref.\cite{wang2023eulerian}.

\begin{figure}
  \centering
  \includegraphics[width=3in]{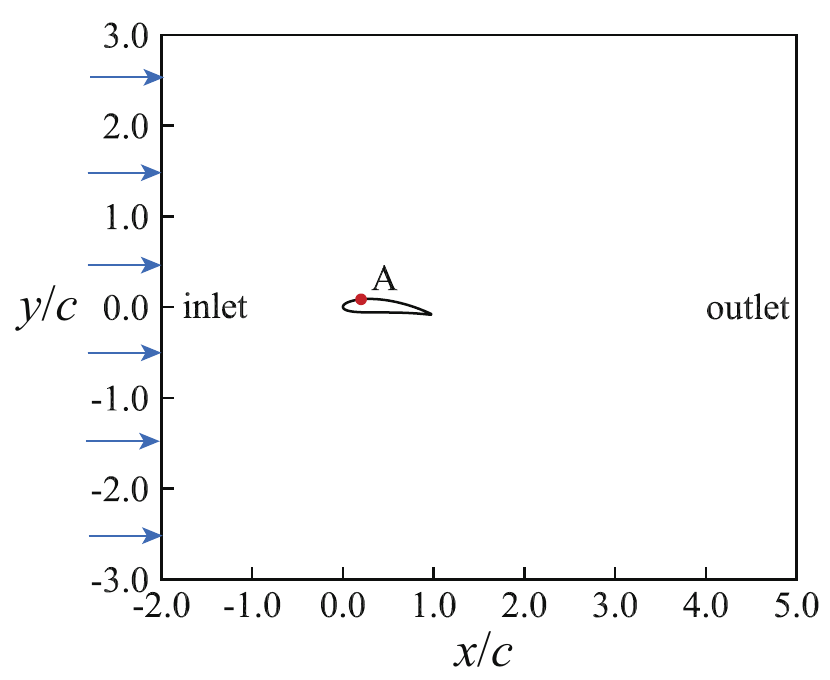}
  \caption{Schematic of NACA 5515 airfoil with AOA $5^\circ$ and computational domain.}
  \label{fig: naca5515-domain}
\end{figure}

\begin{figure}
	\centering
	\begin{subfigure}[b]{0.71\textwidth}
		\centering
		\includegraphics[width=\textwidth]{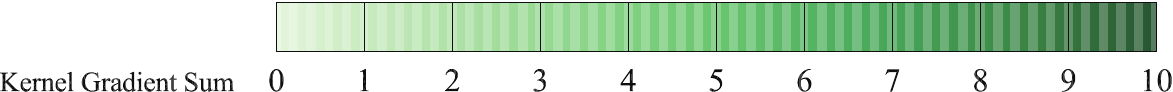}
	\end{subfigure}
	%\newline 
	\begin{subfigure}[b]{0.3\textwidth}
		\centering
		\includegraphics[width=\textwidth]{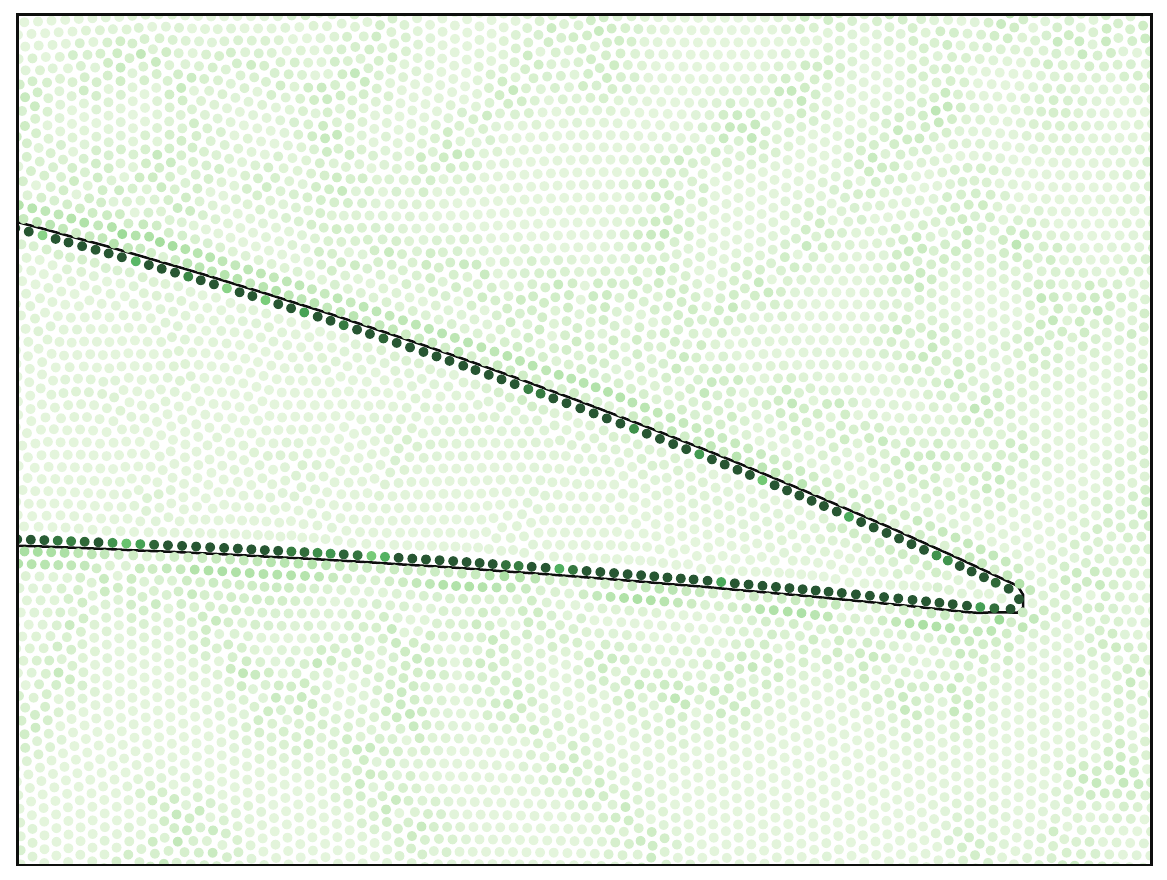}
		\caption{}
		\label{fig: naca5515-complex-consistency}
	\end{subfigure}
	%\newline 
	\begin{subfigure}[b]{0.3\textwidth}
		\centering
		\includegraphics[width=\textwidth]{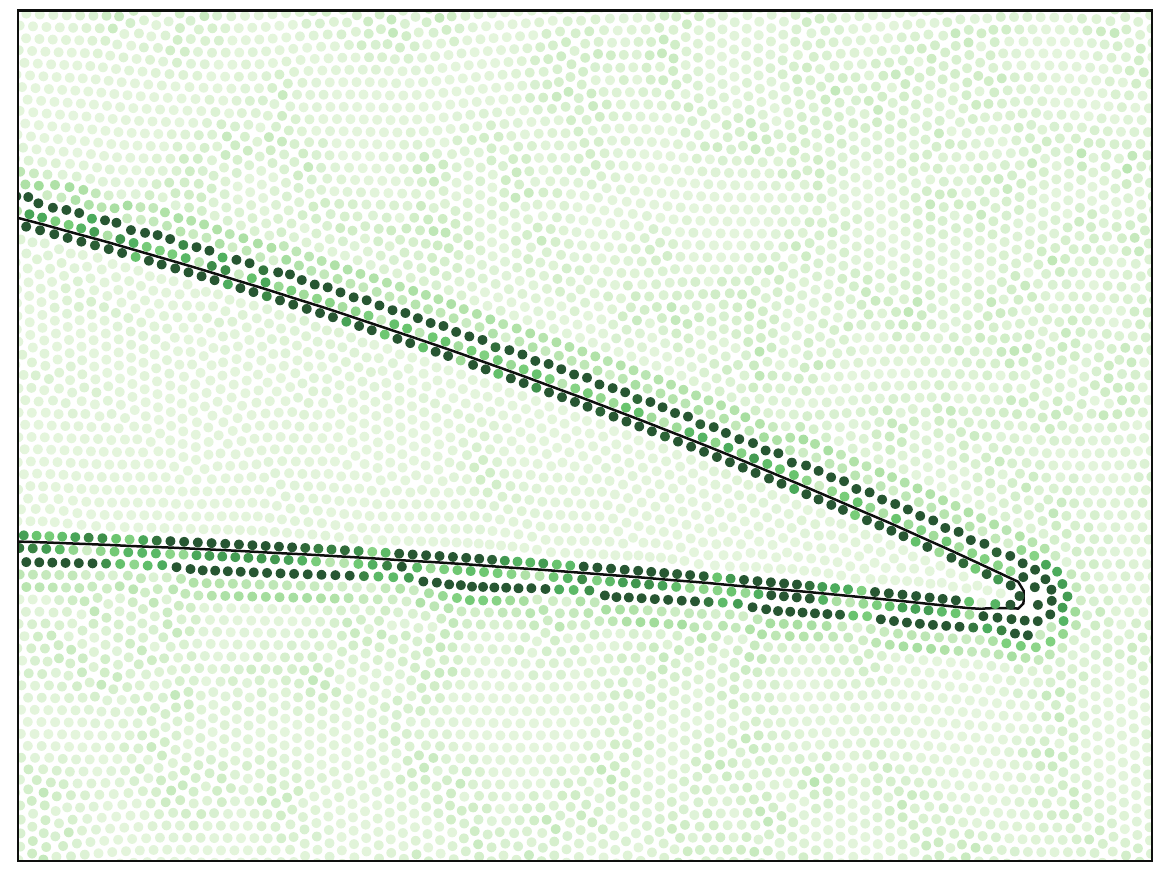}
		\caption{}
		\label{fig: naca5515-separate-consistency}
	\end{subfigure}
        %\newline 
	\begin{subfigure}[b]{0.3\textwidth}
		\centering
		\includegraphics[width=\textwidth]{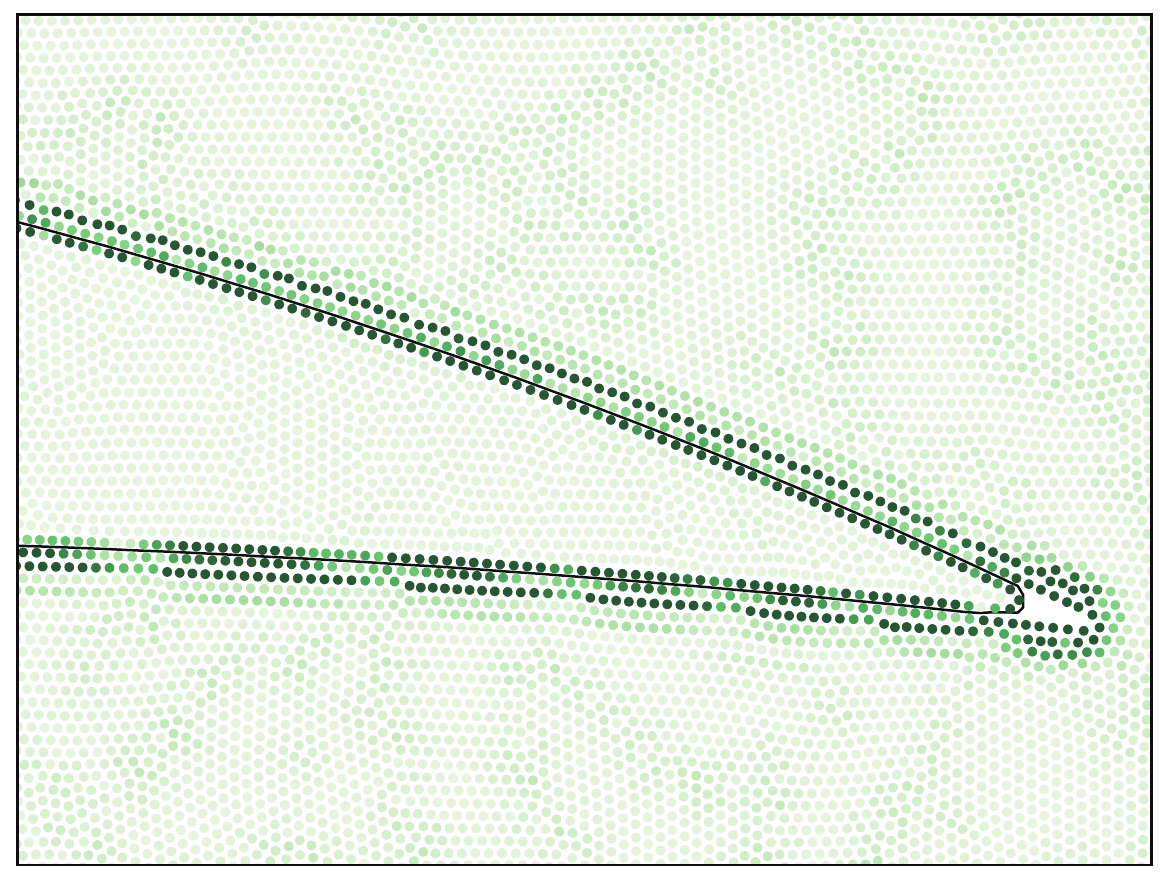}
		\caption{}
		\label{fig: naca5515-separate-gap-consistency}
	\end{subfigure}
	\caption{Kernel gradient summation for the NACA 5515 airfoil with $\Delta x = c/250$ 
 by (a) complex relaxation, (b) separate relaxation and (c) separate relaxation without Boolean 
 operations on geometry. }
	\label{fig: naca5515-consistency}
\end{figure}

\begin{figure}
  \centering
  \includegraphics[width=3in]{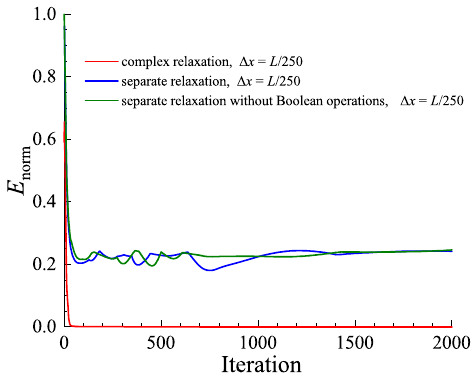}
  \caption{Normalized particle kinetic energy during the physics-driven relaxation process 
  for the NACA 5515 airfoil.}
  \label{fig: naca5515-kinetic-energy}
\end{figure}

The schematic of the NACA 5515 airfoil as well as the computational domain display in 
Fig.\ref{fig: naca5515-domain}, with an angle of attack (AOA) set at $5^\circ$ and a Reynolds number 
($Re$) of 420. Point A ($0.2c$, $0.086c$) is an observation point to monitor the time-dependent 
pressure coefficients, verifying the numerical convergence of the simulation. 
Fig.\ref{fig: naca5515-complex-consistency} and Fig.\ref{fig: naca5515-separate-consistency} display 
the particle distribution near the trailing edge at a resolution of $\Delta x = c/250$ obtained by 
complex relaxation and separate relaxation respectively, where $c$ denotes the chord length of the airfoil. 
Note that in the particle distribution achieved by separate relaxation without Boolean operations,
as also observed in Section \ref{section: 30P30N}, there is a noticeable gap at 
the trailing edge of the airfoil in Fig.\ref{fig: naca5515-separate-gap-consistency} due to the different
geometric parse between fluid and solid bodies. 
However, although the void region of particles 
near the trailing edge has been significantly reduced in the particle distribution within the geometries 
based on Boolean operations as shown in Fig.\ref{fig: naca5515-separate-consistency}, its zero-order 
consistency of particles at the boundary of the fluid domain is significantly inferior compared to 
that obtained through complex relaxation. Meanwhile, it can also be seen from the particle kinetic 
energy diagram, as shown in Fig.\ref{fig: naca5515-kinetic-energy}, that the convergence result obtained 
by complex relaxation is obviously better than that obtained by two separate relaxation methods.

\begin{figure}
  \centering
  \includegraphics[width=3in]{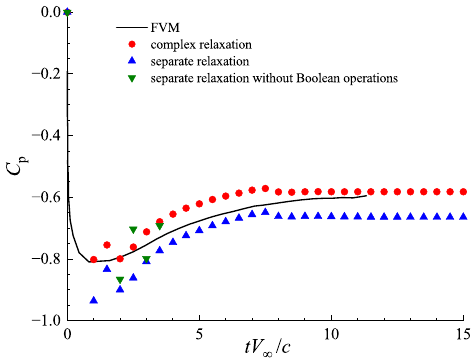}
  \caption{Comparison of time-dependent pressure coefficients at point A among FVM \cite{huang2019kernel}, 
  complex relaxation and separate relaxation.}
  \label{fig: naca-cp-a}
\end{figure}

Fig.\ref{fig: naca-cp-a} presents the time-dependent pressure coefficients $C_p$ at Point A, comparing 
the results from the finite volume method (FVM) as reported in Ref.\cite{huang2019kernel} with Eulerian 
SPH simulations adopting particle generation from complex relaxation and separate relaxation. 
$V_\infty$ is the upstream velocity of the flow. The physical simulation utilizing the particle 
distribution from separate relaxation shows a rapid collapse after only a few time steps. 
At the initial stage of the SPH simulation, there are some fluctuations in the pressure coefficients. 
However, as the simulation progresses, the pressure coefficient from the particle distribution obtained 
via complex relaxation converges much closer to the FVM solution than the result from separate relaxation 
with geometry based on Boolean operations. 

\begin{figure}
  \centering
  \includegraphics[width=3in]{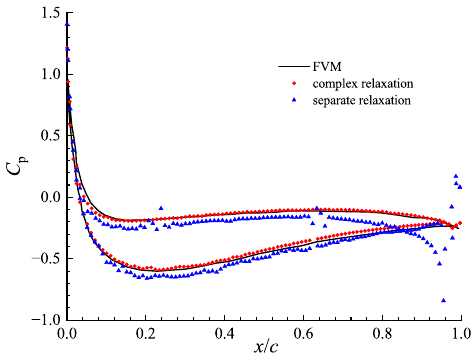}
  \caption{Pressure coefficients on NACA 5515 airfoil surface predicted by FVM \cite{huang2019kernel} 
  and Eulerian SPH with particle distribution achieved by complex relaxation.}
  \label{fig: naca-cp-airfoil}
\end{figure}

Fig.\ref{fig: naca-cp-airfoil} illustrates the pressure coefficients on the surface of the NACA 5515 
airfoil at a non-dimensional time of $tV_\infty/c = 15$, as predicted by Eulerian SPH using a particle 
distribution achieved through complex relaxation and separate relaxation with geometry based on Boolean 
operations. This moment is at the end of the simulation and the system has reached a steady state. 
From this figure, it can be seen that the pressure coefficients obtained from Eulerian SPH with complex 
relaxation change very smoothly along the chordwise of the airfoil and have a close correspondence with 
that from the FVM, even at the trailing edge. However, a small void of particles occurs at the 
trailing edge when the particle arrangement is adopted by separate relaxation as shown in 
Fig.\ref{fig: naca5515-separate-consistency}, although Boolean operations 
ensure no gap problem in the interface during mesh discretization, leading to a bad prediction of pressure 
coefficients in this region. The particle distribution with poor zero-order consistency at the interface by 
separate relaxation also results in a fluctuating change along the airfoil surface. 

\begin{figure}
	\centering
	\begin{subfigure}[b]{0.71\textwidth}
		\centering
		\includegraphics[width=\textwidth]{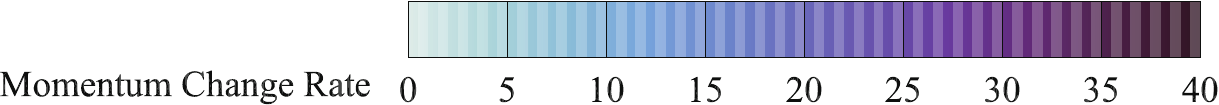}
	\end{subfigure}
	 %\newline 
	\begin{subfigure}[b]{0.3\textwidth}
		\centering
		\includegraphics[width=\textwidth]{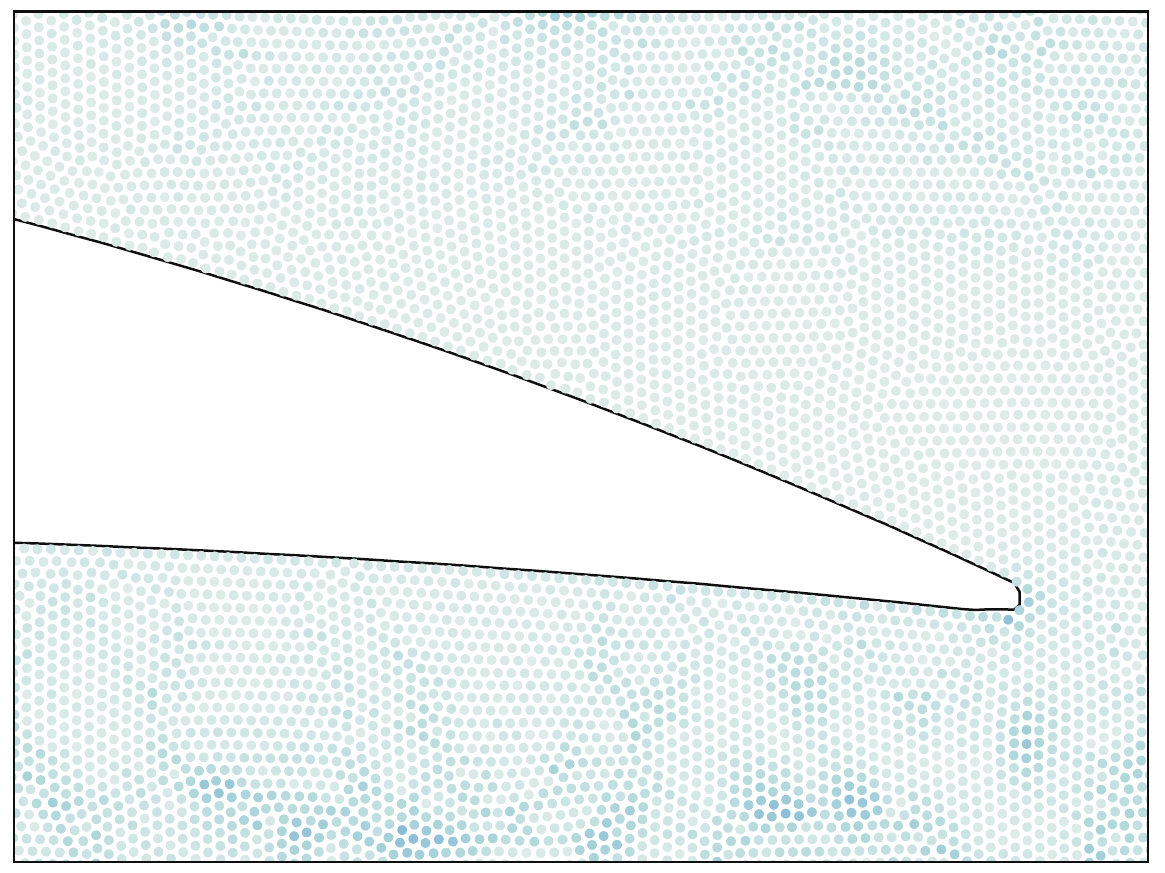}
		\caption{}
		\label{fig: naca5515-complex-mom}
	\end{subfigure}
        %\newline 
	\begin{subfigure}[b]{0.3\textwidth}
		\centering
		\includegraphics[width=\textwidth]{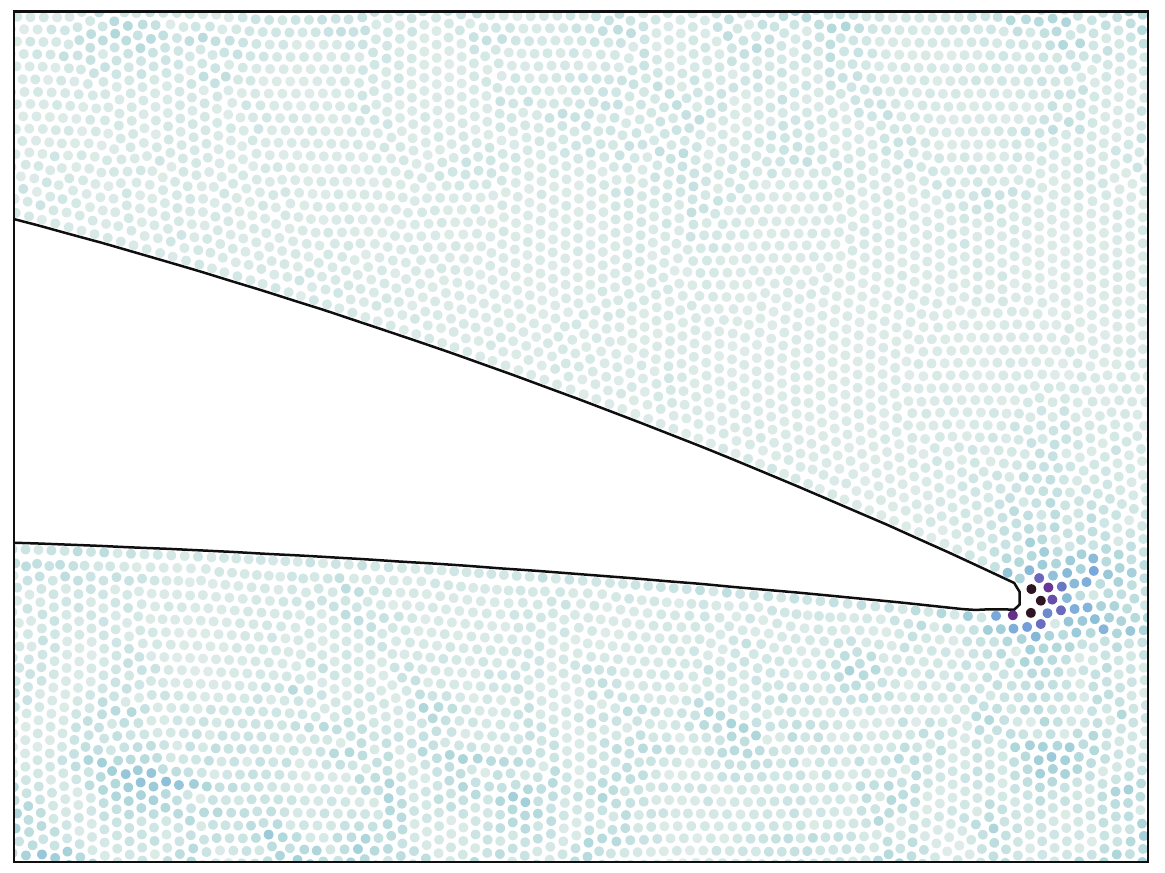}
		\caption{}
		\label{fig: naca5515-separate-mom}
	\end{subfigure}
        %\newline 
	\begin{subfigure}[b]{0.3\textwidth}
		\centering
		\includegraphics[width=\textwidth]{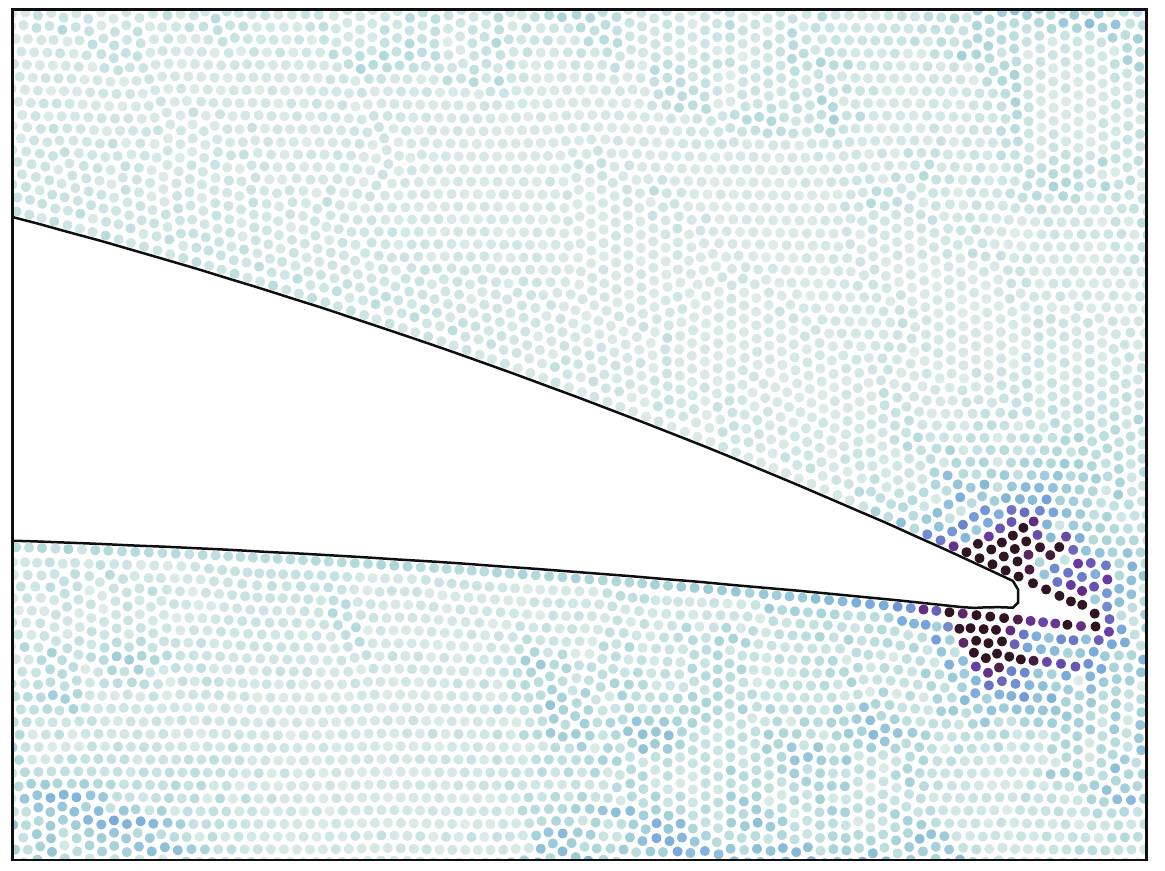}
		\caption{}
		\label{fig: naca5515-separate-gap-mom}
	\end{subfigure}
	\caption{Momentum change rate distribution in the fluid at the moment before the end or collapse, with 
  particles arranged via (a) complex relaxation, (b) separate relaxation and (c) separate relaxation without 
  Boolean operations on geometry.}
	\label{fig: naca5515-momentum-change-rate}
\end{figure}

To understand the reason why physical simulations with particles arranged via separate relaxation 
without Boolean operations collapse after 
only a few time steps, the following analysis has been carried out.
The poor consistency leads to a sudden change in the momentum change rate calculated by the kernel gradient at 
the trailing edge, as shown in Fig.\ref{fig: naca5515-separate-gap-mom} compared with that uniform 
distribution simulated by particles arranged from complex relaxation in 
Fig.\ref{fig: naca5515-complex-mom}. Additionally, the momentum change rate at the trailing 
edge obtained by separate relaxation with geometry based on Boolean operations in 
Fig.\ref{fig: naca5515-separate-mom} also has a high-value region. 

\begin{figure}
  \centering
  \includegraphics[width=3in]{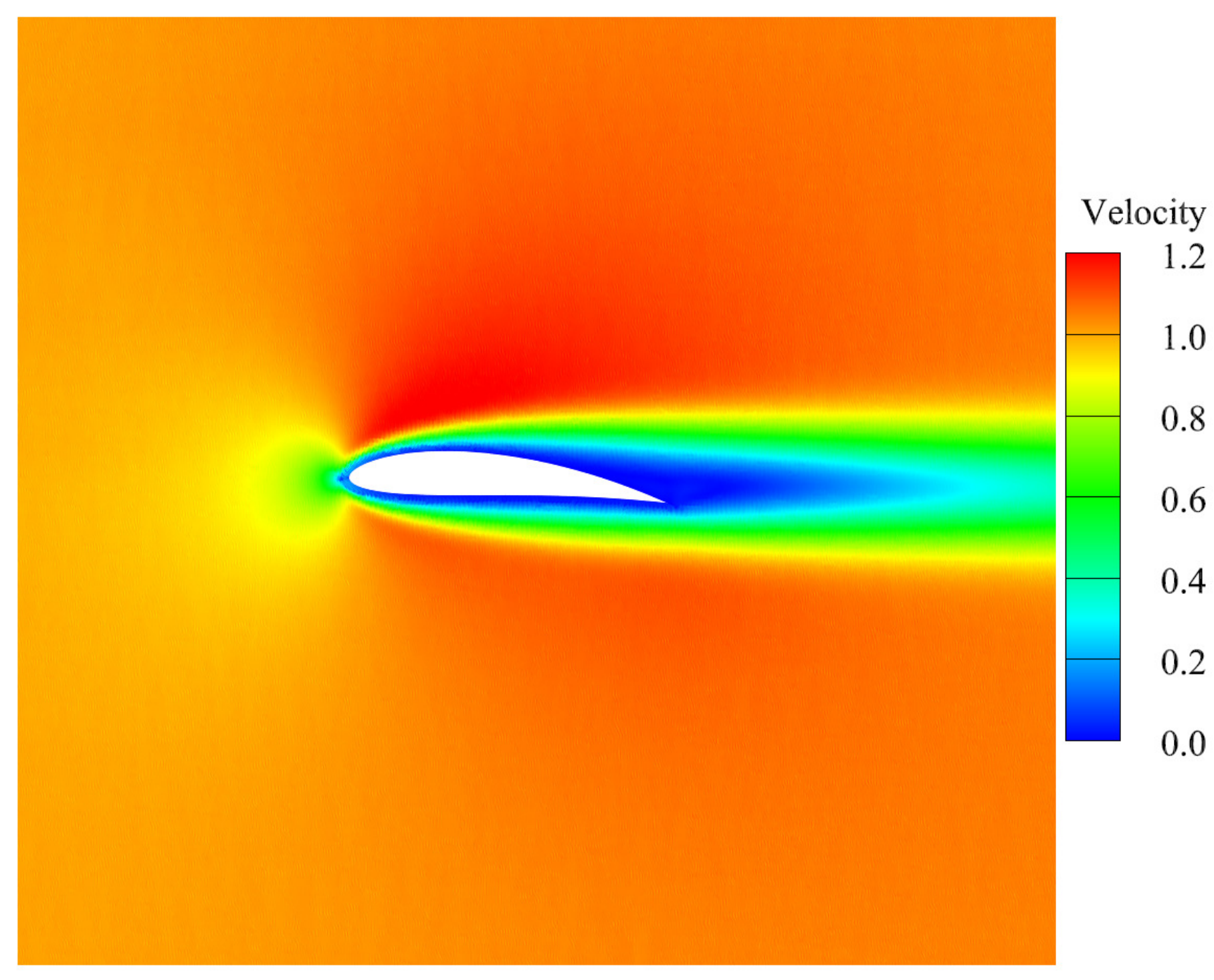}
  \caption{Velocity predicted by Eulerian SPH with particle distribution achieved by complex 
  relaxation.}
  \label{fig: naca-velocity}
\end{figure}

Furthermore, Fig.\ref{fig: naca-velocity} presents the velocity contour from complex relaxation 
at $tV_\infty/c = 15$. This velocity field exhibits a smooth and uniform distribution, further validating the 
effectiveness of the complex relaxation method in distributing particles in SPH simulations within multi-body systems. 
This approach is important for enhancing the quality of particle distribution near interfaces, which contributes to 
achieving accurate and reliable numerical results.

%%%%%%%%%%%%%%%%%%%%%%%%%%%%%%%%%%%%%%%%%%%%%%%%%%%%%%%%%%%%%
%
% Section
%
%%%%%%%%%%%%%%%%%%%%%%%%%%%%%%%%%%%%%%%%%%%%%%%%%%%%%%%%%%%%%
\section{Conclusions} \label{section: conclusions}

In this paper, we introduce a particle relaxation method for multi-body systems. This study mainly 
includes the following procedures: Initially, geometry construction adopts the binary shape approach, 
ensuring shared shape boundary at the geometric interface to eliminate gaps resulting from disparate 
discrete starting points. Subsequently, particles of the inner solid body undergo relaxation under constant 
background pressure, with additional adjustments made for near-surface particles through the surface 
bounding and static confinement corrections. At the same time, the outer fluid body experiences complex 
relaxation, wherein particles near the interface utilize information from both their internal 
configuration as well as the contact solid body, ensuring that a higher zero-order consistency can 
be achieved at the interface region.

The complex relaxation method does not require the pre-arrangement of boundary nodes on the geometric 
surfaces or the application of ghost particles. Through a series of benchmark cases in both two and 
three dimensions, we demonstrate that this method is capable of achieving uniform particle distribution 
in multi-body systems, along with good zero-order consistency. Furthermore, for Eulerian SPH, which 
has a higher demand on the initial particle distribution, the significant contribution of the complex 
relaxation method to the feasibility, reliability and accuracy of physical simulations is exemplified 
by the case study of flow around an airfoil. 

%%%%%%%%%%%%%%%%%%%%%%%%%%%%%%%%%%%%%%%%%%%%%%%%%%%%%%%%%%%%%
%
% Section
%
%%%%%%%%%%%%%%%%%%%%%%%%%%%%%%%%%%%%%%%%%%%%%%%%%%%%%%%%%%%%%
\section*{Acknowledgement}
C.X. Zhao is fully supported by the China Scholarship Council (CSC) (No:202206280028).
Y.C. Yu is fully supported by the China Scholarship Council (CSC) (No:201806120023).
X.Y. Hu would like to express their gratitude to Deutsche Forschungsgemeinschaft for their sponsorship 
of this research under grant number DFG HU1572/10-1 and DFG HU1527/12-1. 
%%%%%%%%%%%%%%%%%%%%%%%%%%%%%%%%%%%%%%%%%%%%%%%%%%%%%%%%%%%%%
%
% Section
%
%%%%%%%%%%%%%%%%%%%%%%%%%%%%%%%%%%%%%%%%%%%%%%%%%%%%%%%%%%%%%
\section*{CRediT authorship contribution statement}
{\bfseries  Chenxi Zhao:} Investigation, Methodology, Visualization, Validation, Formal analysis, 
Writing - original draft, Writing - review \& editing;
{\bfseries  Yongchuan Yu:} Investigation, Methodology, Formal analysis, Writing - review \& editing; 
{\bfseries  Xiangyu Hu:} Investigation, Methodology, Supervision, Writing - review \& editing;
{\bfseries  Oskar J. Haidn:} Investigation, Supervision, Writing - review \& editing.
%%%%%%%%%%%%%%%%%%%%%%%%%%%%%%%%%%%%%%%%%%%%%%%%%%%%%%%%%%%%%
%
% Section
%
%%%%%%%%%%%%%%%%%%%%%%%%%%%%%%%%%%%%%%%%%%%%%%%%%%%%%%%%%%%%%
\section*{Declaration of competing interest }
The authors declare that they have no known competing financial interests 
or personal relationships that could have appeared to influence the work reported in this paper.
%%%%%%%%%%%%%%%%%%%%%%%%%%%%%%%%%%%%%%%%%%%%%%%%%%%%%%%%%%%%%
%
% Section
%
%%%%%%%%%%%%%%%%%%%%%%%%%%%%%%%%%%%%%%%%%%%%%%%%%%%%%%%%%%%%%
\section*{Data availability}
The code is open source on \href{https://www.sphinxsys.org}{https://www.sphinxsys.org}.
%%%%%%%%%%%%%%%%%%%%%%%%%%%%%%%%%%%%%%%%%%%%%%%%%%%%%%%%%%%%%
%
% Section
%
%%%%%%%%%%%%%%%%%%%%%%%%%%%%%%%%%%%%%%%%%%%%%%%%%%%%%%%%%%%%%
\clearpage
%\section*{References}
\bibliography{reference}
%%%%%%%%%%%%%%%%%%%%%%%%%%%%%%%%%%%%%%%%%%%%%%%%%%%%%%%%%%%%%
%
% Document end
%
%%%%%%%%%%%%%%%%%%%%%%%%%%%%%%%%%%%%%%%%%%%%%%%%%%%%%%%%%%%%%
\end{document}